\documentclass[12pt]{article}
\usepackage{psfig,epsf}
\usepackage{a4wide,graphicx}
\usepackage{cite}

\newcommand{\Ima}{\textrm{Im}}
\newcommand{\Rea}{\textrm{Re}}
\newcommand{\mev}{\textrm{ MeV}}
\newcommand{\be}{\begin{equation}}
\newcommand{\ee}{\end{equation}}
\newcommand{\ba}{\begin{eqnarray}}
\newcommand{\ea}{\end{eqnarray}}

\begin{document}

\title{Low lying 
axial-vector mesons as dynamically generated resonances}

\author{
L.~Roca, E.~Oset and J.~Singh\\
{\small Departamento de F\'{\i}sica Te\'orica and IFIC,
Centro Mixto Universidad de Valencia-CSIC,} \\ 
{\small Institutos de
Investigaci\'on de Paterna, Aptdo. 22085, 46071 Valencia, Spain}\\ 
}

\date{\today}

\maketitle

 \begin{abstract} 
 
We make a theoretical study of the s-wave interaction of  the
nonet of vector mesons with the octet of pseudoscalar mesons
starting from a chiral invariant Lagrangian and implementing
unitarity in coupled channels. By looking for poles in the
unphysical Riemann sheets of the unitarized scattering
amplitudes, we get two octets and one singlet of axial-vector
dynamically generated resonances. The poles found can be
associated to most of the low lying axial-vector resonances
quoted in the Particle Data Book: $b_1(1235)$, $h_1(1170)$,
$h_1(1380)$, $a_1(1260)$, $f_1(1285)$ and  two poles
to the $K_1(1270)$ resonance. We evaluate the
couplings of the resonances to the $VP$ states and the partial
decay widths in order to reinforce the arguments in the
discussion.

\end{abstract}

\section{Introduction}

 The realization that QCD at low energies can be studied by
means of effective chiral Lagrangians using fields
associated to the observables mesons and baryons
\cite{Weinberg:1978kz,gasser,ulf,ecker} has brought a
substantial progress in hadronic physics by means of chiral
perturbation theory. A further useful step in this direction
has been given with the introduction of unitary techniques
which have allowed to extend to higher energies the
predictions of chiral perturbation theory, as well as
to tackle new problems barred to a pure perturbative
expansion \cite{Dobado:1996ps,Kaiser:1995cy,Kaiser:1996js,
ollernpa,ollerIAM,ollerND}. The unitary extensions of chiral
perturbation theory, $U\chi PT$, have brought new light in the
study of the meson meson and meson baryon  interaction and
have shown that some well known resonances qualify as
dynamically generated, or in simpler words, they  are
quasibound states of meson meson or meson baryon. This is
the case of the low lying scalar mesons $\sigma$,
$f_0(980)$, $a_0(980)$, $\kappa(900)$
\cite{ollernpa,ollerIAM,ollerND,Kaiser:1998fi,Markushin:2000fa},
which appear from the interaction of pseudoscalar mesons.
Another case of successful
application of these chiral unitary techniques is the
interaction of  mesons with  baryons 
\cite{Kaiser:1995cy,Kaiser:1996js,kaon,Nacher:1999vg,
Oller:2000fj,Inoue:2001ip}
showing that the $\Lambda (1405)$ and the $N^*(1535)$ were
dynamically generated resonances. A more systematic study of
these latter interaction   has shown that there are two
octets and one singlet of resonances from the interaction of
the octet of pseudoscalar mesons with the octet of stable
baryons \cite{Jido:2003cb,nieves}.  Work
along  these lines has continued by studying the interaction
of the octet of pseudoscalar mesons with the decuplet of
baryons \cite{kolo,sourav} which has also led to the
generation of many known resonances, like the $N^*(1520)$
and the $\Lambda(1520)$. 

The studies of the meson meson and meson baryon interaction
along these lines have also shown that some mesons or baryons are
not dynamically generated, they are not consequence of the
interaction between the meson or baryon components and they
qualify better as genuine, or preexistent states, a word that
can be substantiated as that they would remain in the limit of
large $N_c$ where the loops of intermediate states vanish.  This
is the case of the vector mesons of the $\rho$ octet
\cite{ollerND,Pelaez:2003dy}.

A suggestive step forward in the direction of studying the meson
meson interaction along these lines would be to study the
interaction of the octet of pseudoscalar mesons with the octet
of vector mesons (nonet including $\phi$ and $\omega$ with the
standard mixing).  Work in this direction has already been done
in \cite{lutzaxials} and, interestingly,  most of the low lying
axial-vector mesons appear as dynamically generated resonances.
If this were the case, this would have many practical
as well as conceptual repercussions and, hence, extra efforts to
corroborate these findings, looking also for uncertainties, are
called for. In \cite{lutzaxials} the search for resonances was
done by looking at the speed plots of the physical amplitudes.
A search for poles of the amplitudes in the unphysical Riemann
sheets in the complex plane is a more powerful tool to
investigate resonances an a thorough study along these lines is
called for. The present work goes in this last direction and we
have done a thorough work investigating the following points:
1)  The poles of the amplitudes have been systematically
searched for in the complex energy plane following the
trajectories in terms of an SU(3) breaking parameter. This has
allowed us to make an SU(3) study of the problem as well as the
effects of SU(3) symmetry breaking, very useful to understand
the meaning of the poles and their number. 2) The study of the
amplitudes around the poles has also allowed us to determine the
coupling of the resonances to the different coupled channels and
from there the partial decay widths into the different
channels.  This study has been useful to make an association of
the resonances found with those of the Particle Data Group (PDG)
\cite{pdg}, (see Table~\ref{tab:octets}). 3) We have changed the
association of the poles to the known resonances in some cases
with respect to \cite{lutzaxials}, in particular in the case of
the  $K_1(1270)$ resonance, which we claim comes from two
distinct poles with very different properties. This has some
practical consequences in the partial decay widths and the 
dominance of certain channels in different reactions, which
would help clarifying the puzzle of these resonances. 4) At the
same time, and in order to help new researchers in the field, we
have substantially simplified  the formalism of
\cite{lutzaxials}, making it more amenable and transparent.

The results obtained here support the bulk of
the claims made in \cite{lutzaxials},
but the larger amount of information on these resonance obtained
here brings new elements that set on firmer grounds the
association of these resonances with the experimental ones.
This substantiates the
idea that the low lying axial-vector resonances are
dynamically generated, with the exception of the higher
mass ones, the $f_1(1420)$ and the $K_1(1400)$, which
do not fit easily in our scheme, while at the same time we
suggest that the $K_1(1270)$ corresponds actually to two
poles, which would have many experimental repercussions.

\begin{table}[htpb]
\begin{center}
\begin{tabular}{|c||c|c|cc|}\hline &&& \\[-4.4mm]
$J^{PC}$ &$I=1$  &$I=0$ & $I=1/2$ & \\ \hline \hline  &&& \\[-4.4mm]
 $1^{+-}$  & $b_1(1235)$  & $h_1(1170)$, $h_1(1380)$
  &  $K_{1B}$ & \\ \cline{1-4} &&&& \\ [-7.0mm]  &&&&
  $K_1(1270),\,K_1(1400)$ \\[-2.2mm]
 $1^{++}$  & $a_1(1260)$  & $f_1(1285)$, $f_1(1420)$
    &$K_{1A}$ &\\ \hline     
\end{tabular}
\end{center}
\caption{$1^+$ resonances in the Particle Data Book \cite{pdg}.
The $K_1(1270)$ and $K_1(1400)$ are assumed to be a mixture of
$K_{1A}$ and $K_{1B}$.}
\label{tab:octets}
\end{table}

\section{Pseudoscalar-vector meson interaction}

\subsection{Tree level potential
\label{subsec:treeV}} 

There is not a unique formulation to treat the vector mesons in
an effective theory. The ambiguity comes from the freedom  in
considering the nature of the vector meson to be or not  a
gauge boson of a certain symmetry, and from the election of their
transformation properties under a certain realization of chiral
symmetry. For a review see, for example, Ref.~\cite{birse}.
Despite the differences between the treatments
of vector mesons,
the equivalence between the different approaches can be shown
at lowest order \cite{birse}. Considering the vector mesons as
fields transforming homogeneously under the nonlinear
realization of chiral symmetry, the interaction of two vector
and two pseudoscalar mesons at lowest order in the pseudoscalar
fields can be obtained from the following interaction Lagrangian 
\cite{birse}\footnote{
Note the different factor $-1/4$ instead of $-1/2$ in 
\cite{birse} to agree with our normalization of the fields.}

\begin{equation}
{\cal L}_{I}=-\frac{1}{4}Tr\left\{\left(\nabla_\mu V_\nu-
\nabla_\nu V_\mu\right)
\left(\nabla^\mu V^\nu-\nabla^\nu V^\mu\right)\right\},
\label{eq:LBirse}
\end{equation}
\noindent
where $Tr$ means $SU(3)$ trace and $\nabla_\mu$ is the covariant
derivative defined as 
\begin{equation}
\nabla_\mu V_\nu=\partial_\mu V_\nu+[\Gamma_\mu , V_\nu],
\end{equation}
\noindent
where $[,]$ stands for commutator and $\Gamma_\mu$ is the
vector current
\begin{equation}
\Gamma_\mu=\frac{1}{2} (u^\dagger\partial_\mu u
+u\partial_\mu u^\dagger)
\end{equation}
\noindent
with 
\begin{equation}
u^2=U=e^{i\frac{\sqrt{2}}{f}P}.
\label{eq:ufields}
\end{equation}
\noindent
In the previous equations $f=92\mev$ is the pion decay constant
and $P$ and $V$ are the $SU(3)$ matrices containing the octet of
pseudoscalar and the nonet of vector mesons respectively:

\be
P \equiv 
\left(\begin{array}{ccc}
 \frac{1}{\sqrt{2}} \pi^0 + \frac{1}{\sqrt{6}}\eta_8 
 & \pi^+ & K^+\\
\pi^-&- \frac{1}{\sqrt{2}} \pi^0 + \frac{1}{\sqrt{6}}\eta_8 
& K^0\\
K^-& \bar{K}^0 & -\frac{2}{\sqrt{6}}\eta_8
\end{array}
\right) 
, \,\,
V_\mu \equiv  \left(\begin{array}{ccc} 
\frac{1}{\sqrt{2}} \rho^0 + \frac{1}{\sqrt{2}}\omega 
 & \rho^+ & K^{*+}\\
\rho^-& - \frac{1}{\sqrt{2}} \rho^0 + \frac{1}{\sqrt{2}}\omega 
& K^{*0}\\
K^{*-}& \bar{K}^{*0} & \phi
\end{array}
\right)_{\mu} .
\label{eq:PVmatrices}
\ee

The Lagrangian of Eq.~(\ref{eq:LBirse}) is invariant under the
chiral transformations $SU(3)_L\otimes SU(3)_R$, since
$\nabla_\mu V_\nu$ transforms as \cite{egpdr}
\be
\nabla_\mu V_\nu \to h\nabla_\mu V_\nu h^\dagger.
\ee
We are interested in the two-vector--two-pseudoscalar amplitudes.
Hence, expanding the Lagrangian  of Eq.~(\ref{eq:LBirse}) up to
two pseudoscalar meson fields we find
\begin{equation}
{\cal L}_{VVPP}=-\frac{1}{4f^2}
Tr\left([V^{\mu},\partial^{\nu}V_{\mu}]
        [P,\partial_{\nu}P]\right),
\label{eq:L}
\end{equation}
\noindent
which would account for the Weinberg-Tomozawa interaction for the
$VP\to VP$ process 
\cite{lutzaxials,birse,Weinberg:1966kf,Tomozawa:1966jm}.

Note that in Eq.~(\ref{eq:PVmatrices})
in the pseudoscalar octet we are only considering the
$\eta_8\equiv\eta$ and not the $\eta'$. The inclusion of the
$\eta'$ effects in strong interactions can be accommodated in the
higher order Lagrangians \cite{gassleut}. 
Since the meson decay constant, $f$, is different for different
mesons, one could also use different values of $f$ for the
different pseudoscalars, as done in \cite{Inoue:2001ip,kaon}. We
shall comment on the results obtained when we take this into
account.

In the vector meson multiplet we have assumed ideal 
$\omega_1-\omega_8$ mixing:
\be
\phi=\omega_1/\sqrt3-\omega_8\sqrt{2/3},\quad
\omega=\omega_1\sqrt{2/3}+\omega_8/\sqrt{3}.
\label{eq:phiw}
\ee
 (Throughout the work we will use the phase convention
  $|\pi^+>=-|1+1>$, 
 $|\rho^+>=-|1+1>$, $|K^{-}>=-|1/2-1/2>$ and 
 $|K^{*-}>=-|1/2-1/2>$ corresponding $|I I_3>$ isospin states).

From the Lagrangian of Eq.~(\ref{eq:L})
one obtains the full amplitude which
 we deduce in Appendix~I,
where we also make the projection over s-wave, which leads to

\begin{equation}
V_{ij}(s)=-\frac{\epsilon\cdot\epsilon'}{8f^2} C_{ij}
\left[3s-(M^2+m^2+M'^2+m'^2)
-\frac{1}{s}(M^2-m^2)(M'^2-m'^2)\right],
\label{eq:Vtree}
\end{equation}
\noindent
where $\epsilon$($\epsilon'$) stands for the polarization 
four-vector of the incoming(outgoing) vector meson. The masses
$M(M')$,  $m(m')$ correspond to the initial(final) vector mesons
and initial(final) pseudoscalar mesons respectively, and we use an
averaged value for each isospin multiplet. The indices $i$ and $j$
represent the initial and final $VP$ states respectively.

We can identify the $VP$ channels by its strangeness ($S$)  and
isospin ($I$), $(S,I)=(1,1/2)$, $(0,0)$ and $(0,1)$. There are
other possible $(S,I)$ combinations but since, advancing some
results, we will not find poles there, we will not consider 
them in the discussion. For the $(0,0)$ channel the allowed
isospin channels are $\bar K^*K$, $\phi\eta$, $\omega\eta$,
$\rho\pi$ and $K^*\bar K$, for the $(0,1)$ channels they are
$\bar K^*K$, $\phi\pi$, $\omega\pi$, $\rho\eta$, $\rho\pi$ and
$K^*\bar K$, and for $(1,1/2)$ we have $\phi K$, $\omega K$,
$\rho K$, $K^*\eta$ and $K^*\pi$. Note that for the $(0,0)$ and
$(0,1)$ cases, the isospin states have well-defined
$G-$parity\footnote{Recall  that the $G-$parity operation can be
defined through its action on an eigenstate of hypercharge
($Y$), isospin  ($I$), and third isospin projection ($I_3$)  as 
$G|Y,I,I_3>=\eta(-1)^{Y/2+I}|-Y,I,I_3>$, with $\eta$ being the
charge conjugation of a neutral non-strange member of the
$SU(3)$ family.} except the $\bar K^*K$ and $K^*\bar K$ states,
but the combinations $1/\sqrt{2}(|\bar K^* K>\pm|K^*\bar K>)$
are actually $G-$parity eigenstates with eigenvalues $\pm$.

In Tables \ref{tab:Cij5}, \ref{tab:Cij6} and \ref{tab:Cij7} we show the $C_{ij}$ coefficients in isospin base
for 
$(S,I)=(1,1/2)$, $(0,0)$ and $(0,1)$, indicating also the
$G-$parity in the $(0,0)$ and $(0,1)$ cases.

\begin{table}[h]
\begin{center}
\begin{tabular}{|c|ccccc|}
\hline
 & $\phi K$ & $\omega K$ & $\rho K$ & $K^* \eta$ & $K^* \pi$ \\	
\hline  &&&&&\\[-4.5mm]
$\phi K$   & $0$ & $0$ & $0$ & $-\sqrt{\frac{3}{2}}$ & $-\sqrt{\frac{3}{2}}$   \\
&&&&&\\[-4.5mm]
$\omega K$ & $0$ & $0$ & $0$ & $\frac{\sqrt{3}}{2}$ & $\frac{\sqrt{3}}{2}$   \\
&&&&&\\[-4.5mm]
$\rho K$   & $0$ & $0$ & $-2$ & $-\frac{3}{2}$  & $\frac{1}{2}$    \\ 
$K^* \eta$ & $-\sqrt{\frac{3}{2}}$ & $\frac{\sqrt{3}}{2}$ & $-\frac{3}{2}$ & $0$ & $0$ \\
&&&&&\\[-4.5mm]
$K^* \pi$  & $-\sqrt{\frac{3}{2}}$ & $\frac{\sqrt{3}}{2}$ & $\frac{1}{2}$ & $0$ & $-2$ \\
\hline
\end{tabular}
\caption{$C_{ij}$ coefficients in isospin base
 for $S=1$, $I=\frac{1}{2}$.}
\label{tab:Cij5}
\end{center}
\end{table}

\begin{table}[h]
\begin{center}
\begin{tabular}{|c|c|ccccc|}
\hline &&&&&&\\[-4.5mm]
$G$& & $\frac{1}{\sqrt{2}}(\bar K^*K+K^*\bar K)$ & $\phi\eta$ & $\omega\eta$ & $\rho\pi$ & $\frac{1}{\sqrt{2}}(\bar K^*K-K^*\bar K)$ \\	
\hline &&&&&&\\[-4.5mm]
$+$&$\frac{1}{\sqrt{2}}(\bar K^*K+K^*\bar K)$   & $-3$ & $0$ & $0$ & $0$ & $0$	\\
$-$&$\phi\eta$ & $0$ & $0$ & $0$ & $0$ & $\sqrt{6}$   \\
$-$&$\omega\eta$ & $0$ & $0$ & $0$ & $0$  & $-\sqrt{3}$      \\ 
$-$&$\rho\pi$ & $0$ & $0$ & $0$ & $-4$ & $\sqrt{3}$ \\
$-$&$\frac{1}{\sqrt{2}}(\bar K^*K-K^*\bar K)$  & $0$ & $\sqrt{6}$ &  $-\sqrt{3}$ & $\sqrt{3}$ & $-3$ \\
\hline
\end{tabular}
\caption{$C_{ij}$ coefficients in isospin base for $S=0$, $I=0$. The first column indicates the $G-$parity.}
\label{tab:Cij6}
\end{center}
\end{table}

\begin{table}[h]
\begin{center}
\begin{tabular}{|c|c|cccccc|}
\hline &&&&&&&\\[-4.5mm]
$G$& & $\frac{1}{\sqrt{2}}(\bar K^*K+K^*\bar K)$ & $\phi\pi$ & $\omega\pi$ & $\rho\eta$ & $\rho\pi$ & $\frac{1}{\sqrt{2}}(\bar K^*K-K^*\bar K)$ \\	
\hline &&&&&&&\\[-4.5mm]
$+$&$\frac{1}{\sqrt{2}}(\bar K^*K+K^*\bar K)$   & $-1$ & $-\sqrt{2}$ & $1$ & $\sqrt{3}$ & $0$ & $0$  \\
$+$&$\phi\pi$ & $-\sqrt{2}$ & $0$ & $0$ & $0$ & $0$ & $0$   \\
$+$&$\omega\pi$ & $1$ & $0$ & $0$ & $0$ & $0$ & $0$	 \\ 
$+$&$\rho\eta$  & $\sqrt{3}$ & $0$ & $0$ & $0$ & $0$& $0$ \\
$-$&$\rho\pi$  & $0$ & $0$ & $0$&$0$ & $-2$& $\sqrt{2}$\\
$-$&$\frac{1}{\sqrt{2}}(\bar K^*K-K^*\bar K)$  & $0$ & $0$ &  $0$ & $0$ & $\sqrt{2}$ &  $-1$ 
\\[-4.7mm] &&&&&&& \\ \hline
\end{tabular}
\caption{$C_{ij}$ coefficients in isospin base for $S=0$, $I=1$. The first column indicates the $G-$parity.}
\label{tab:Cij7}
\end{center}
\end{table}

Let us now discuss an interesting consequence in the
sign and strength of the potential obtained, which can 
give us an indication about whether this Lagrangian can
produce pseudo-bound states with a suitable 
unitarization procedure. The interaction of two octets
gives the following decomposition in irreducible
representations of  $SU(3)$:

\begin{equation}
8\otimes 8=1\oplus 8_a\oplus 8_s \oplus \bar{10} \oplus 10
\oplus 27.
\label{eq:decomp}
\end{equation}
\noindent
Note that although we shall work with a nonet of vector mesons
(including the $\phi$ and $\omega$) the singlet component does
not lead to an interaction term from Eq.~(\ref{eq:L}).

The coefficients $C_{ij}$ of Eq.~(\ref{eq:Vtree}) can be
expressed in any desired base: charge, isospin, SU(3), etc.
Since the $SU(3)$ matrices are given in terms of the  physical
charge fields, the most easy way to express the $C_{ij}$
coefficients is in this base, but with the use of the $SU(3)$
Clebsch-Gordan coefficients it is straightforward
 to obtain the coefficients
$C$ in the $SU(3)$ base, $\alpha$. They give
\begin{equation}
C_{\alpha\beta}=diag\{-6,-3,-3,0,0,2\}
\label{eq:CSU3}
\end{equation}
\noindent
in the order of $1$, $8_a$, $8_s$, $\bar{10}$, $10$ and $27$. 
As we shall see,
a minus sign in a coefficient of Eq.~(\ref{eq:CSU3})
implies an attractive potential, which is
needed to have a bound state. 
Therefore, in the $SU(3)$ limit, we should expect attraction
in the singlet and the two octets, no interaction in the
decuplets and repulsion in the $27-$plet.
 Therefore, a priori,
one could expect, after the unitarization procedure that we will
explain below, two octets and one singlet of dynamically
generated axial-vector ($J^P=1^+$) resonances.
In addition, in the exact $SU(3)$ symmetric case the two
octets would be degenerate.


\subsection{Unitarization procedure}

In the literature several unitarization procedures have been
used to obtain a scattering matrix fulfilling exact
unitarity in coupled channels, like the Inverse Amplitude
Method \cite{Dobado:1996ps,ollerIAM} or the $N/D$ method
\cite{ollerND}. In this latter work the equivalence with the
Bethe-Salpeter equation used in \cite{ollernpa} was established.

\begin{figure}[h]
\begin{center}
\includegraphics[width=0.8\textwidth]{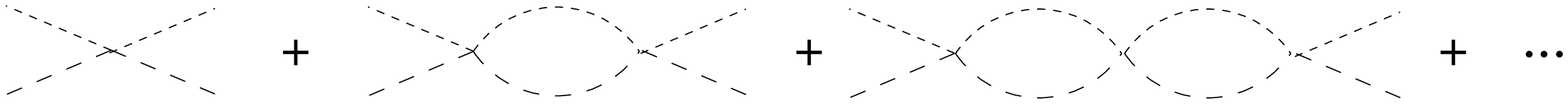}
\caption{Diagrammatic representation of the resummation of loops
in the unitarization procedure.}
\label{fig:bethe}
\end{center}
\end{figure}

In the present work we  make use of the Bethe-Salpeter
equation to resum the
diagrammatic series expressed in Fig.~\ref{fig:bethe}

There are some subtle differences with respect, for instance, to
the pseudoscalar-pseudoscalar case, coming from the polarization
vectors appearing in the potential and the particular form of
the vector meson propagator in the loop. For the sake of clarity
in the exposition we have relegated the explanation to 
Appendix~II. From the reasons explained in  Appendix~II
 we can do
the evaluation of the scattering matrix for transverse
polarization modes of the external vector mesons, which leads to
the following unitarized amplitude:

\begin{equation}
T=[1+V\hat{G}]^{-1}(-V)
\,\vec{\epsilon}\cdot\vec{\epsilon}\,',
\label{eq:bethe}
\end{equation}
\noindent
where $\hat{G}=G(1+\frac{1}{3}\frac{q^2_l}{M_l^2})$
is a diagonal matrix with the $l-$th element, $G_l$, being
the two meson loop function
containing a vector and a pseudoscalar meson:

\begin{equation}
G_{l}(\sqrt{s})= i \, \int \frac{d^4 q}{(2 \pi)^4} \,
\frac{1}{(P-q)^2 - M_l^2 + i \epsilon} \,
 \frac{1}{q^2 - m^2_l + i
\epsilon},
\label{eq:loop}
\end{equation}
\noindent
with $P$ the total incident momentum, which in the center of mass
frame is $(\sqrt{s},0,0,0)$.

The structure of Eq.~(\ref{eq:bethe}) explains why a minus sign in
the $C_{\alpha\beta}$ coefficients of Eq.~(\ref{eq:CSU3})
together with  Eq.~(\ref{eq:Vtree}), implies
attraction, ($\Rea G$ is negative in the region of relevance and
$1+V\hat{G}$ can lead to poles with $V$ positive).

In the dimensional regularization scheme the loop function of 
Eq.~(\ref{eq:loop}) gives
\begin{eqnarray}
G_{l}(\sqrt{s})&=& \frac{1}{16 \pi^2} \left\{ a(\mu) + \ln
\frac{M_l^2}{\mu^2} + \frac{m_l^2-M_l^2 + s}{2s} \ln \frac{m_l^2}{M_l^2} 
\right. \label{propdr} \\ & &  \phantom{\frac{1}{16 \pi^2}} +
\frac{q_l}{\sqrt{s}}
\left[
\ln(s-(M_l^2-m_l^2)+2 q_l\sqrt{s})+
\ln(s+(M_l^2-m_l^2)+2 q_l\sqrt{s}) \right. \nonumber  \\
& & \left. \phantom{\frac{1}{16 \pi^2} +
\frac{q_l}{\sqrt{s}}}
\left. \hspace*{-0.3cm}- \ln(s-(M_l^2-m_l^2)-2 q_l\sqrt{s})-
\ln(s+(M_l^2-m_l^2)-2 q_l\sqrt{s}) -2\pi i \right]
\right\},
\nonumber
\end{eqnarray}
where $\mu$ is the scale of dimensional regularization. Changes
in the  scale are reabsorbed in the subtraction constant
$a(\mu)$, so that the results  remain scale independent. In
Eq.~(\ref{propdr}), $q_l$ denotes the three-momentum of the
vector or pseudoscalar meson in the center of mass frame 
and is given by
\begin{equation}
q_l=\frac{1}{2\sqrt{s}}\sqrt{[s-(M_l+m_l)^2][s-(M_l-m_l)^2]}
\label{eq:q}
\end{equation}
where $M_l$ and $m_l$ are the masses of
the vector and pseudoscalar mesons respectively. For the
evaluation of the loop in the physical, or first, Riemann sheet
one has to take the solution for the square root with
$\Ima(q_l)$ positive. 
Note that in Eq.~(\ref{propdr}) there is an ambiguity in the
imaginary part of the $\ln$
function coming from their multivaluedness.
This ambiguity can be removed, for a generally complex $\sqrt{s}$,
by comparing the result with the one obtained numerically 
by regularizing Eq.~(\ref{eq:loop}) by means of a cutoff
of a natural size, of the order of $1000\mev$. By doing this,
we have checked that the prescription for the $\ln$ which gives a
result in accordance with the cutoff method is to use the $\ln$
with its argument defined with a phase from $-\pi$ to $\pi$ with
a cut in the negative real axis.
On the other hand, this comparison with the cutoff method
allows us also to determine the subtraction constant which turn
out to be of $\sim -1.85$.

In \cite{lutzaxials}
a different regularization procedure is used by choosing  the $G$
function to vanish for $\sqrt{s}$ equal to the mass of the vector
meson. A similar choice, making the $G$ function vanish at the mass
of the baryon is shown to lead to realistic results in the meson
baryon interaction case \cite{nieves}. In the present 
work we will use
both approaches in order to have an idea of the theoretical
uncertainties in the results.

On the other hand, note that we do not include the width of the
vector mesons in their propagators in Eq.~(\ref{eq:loop}). We shall
take this into account in a coming section.


\subsection{Comparison to perturbative expansion}

This unitarization procedure can be understood as an analytical
extrapolation of perturbation theory  to higher energies in the
same way as is done for ordinary chiral perturbation theory with
pseudoscalar meson-meson or meson-baryon interaction 
\cite{ollernpa,Kaiser:1998fi,kaon,Oller:2000fj}.
Perturbation theory  with the Lagrangian of Eq.~(\ref{eq:L})
would proceed like
ordinary chiral perturbation theory for pseudoscalar 
meson-meson and meson-baryon interaction, with loop divergences
canceled with higher order Lagrangians (see \cite{birse}) which
are ordered in terms of increasing number of derivatives in the
fields. The expansion appears as a power series
 of the momentum over one
scale (expansion parameter), $\Lambda_\chi\sim 4\pi f\simeq
1.2\textrm{ GeV }$ (in the chiral limit of pseudoscalar masses
going to zero). At one loop level one would need the next order
Lagrangian to reabsorb the loop divergences and one would have
the direct s-channel loop  as well as the crossed loop term. The
unitary amplitude should match the perturbative expansion at low
energies, but our procedure only provides the s-channel loop and
furthermore does not use information of higher order Lagrangians.
The philosophy behind this is that the contribution of crossed
loops generates a smooth energy dependence in the energy region
of our concern and can be reabsorb in the subtraction constant,
$a$, of the loop function $G$ \cite{ollerND}. Similarly, one is
also assuming that the
one loop calculation with
 the lowest order Lagrangian, with the use of a $G$
function with natural size subtraction constant, can account for
the effect of the higher order Lagrangians. This is what
characterizes the dynamically generated resonances, in contrast
to cases like the $\rho$ meson  (a genuine resonance of basically
two constituent quarks) which require the explicit use of a higher
order Lagrangian \cite{ollerND}.

In order to illustrate the previous discussion, we shall now
compare results for the amplitudes obtained with the unitarity
amplitude of Eq.~(\ref{eq:bethe}) and its perturbative expansion
up to two loops. We choose as an example the $K^*\pi\to K^*\pi$ in
$(S,I)=(1,1/2)$ and $\omega\pi\to\omega\pi$, $(S,I)=(0,1)$. 
\begin{figure}[h]
\begin{center}
\includegraphics[width=0.7\textwidth]{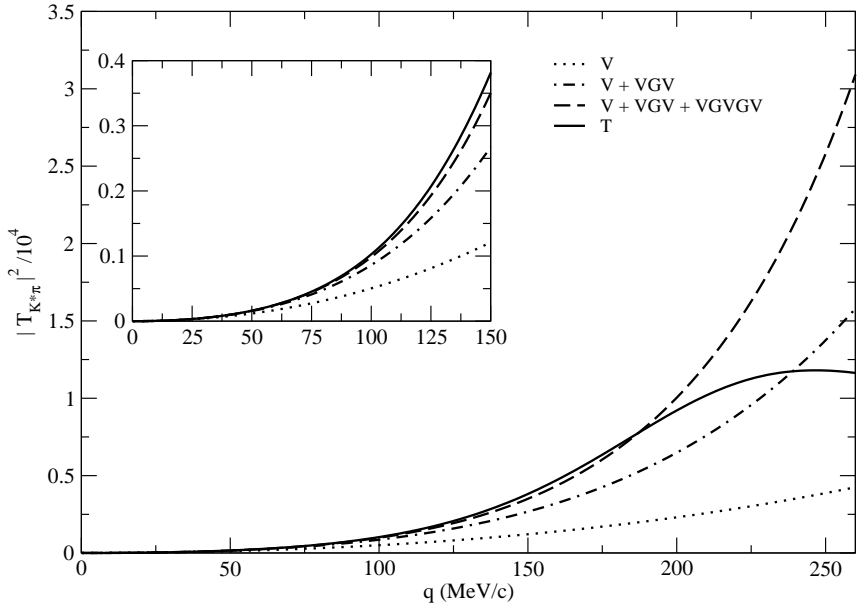}
\caption{Comparison of the perturbative expansion to the
unitarized amplitude for $K^*\pi\to K^*\pi$ 
with only one channel ($K^*\pi$) and in the chiral
 limit ($m_\pi=0$). All the lines
represent the modulus square. Dotted line:
tree level potential $(-V)$; dashed-dotted
 line: expansion of $T$ at one
loop, $(-V)+(-V)G(-V)$; dashed line: expansion of $T$ at two
loops, $(-V)+(-V)G(-V)+(-V)G(-V)G(-V)$; solid line:
unitarized amplitude, $T$. 
}
\label{fig:perturb1}
\end{center}
\end{figure}
First we show, in Fig.~\ref{fig:perturb1},
 results for the $K^*\pi\to K^*\pi$ amplitude using
only one channel ($K^*\pi$) and in the chiral limit ($m_\pi=0$).
In Fig.~\ref{fig:perturb1} we can see the modulus squared of the
amplitudes calculated with the approximations $(-V)$, 
$(-V)+(-V)G(-V)$, $(-V)+(-V)G(-V)+(-V)G(-V)G(-V)$ and the
unitary amplitude. We can see that for low momenta there is a
nice convergence of the perturbative series to the unitary
result up to about $150\mev/c$. However, the unitary amplitude
has a resonant structure with a peak around  $250\mev/c$ where
the perturbative expansion is seen to fail drastically. Note
that the lowest order $(-V)$ (in the chiral limit) is of order
$q$, as can be seen from Eq.~(\ref{eq:LBirse}) or expanding
Eq.~(\ref{eq:L}).

\begin{figure}[h]
\begin{center}
\includegraphics[width=0.7\textwidth]{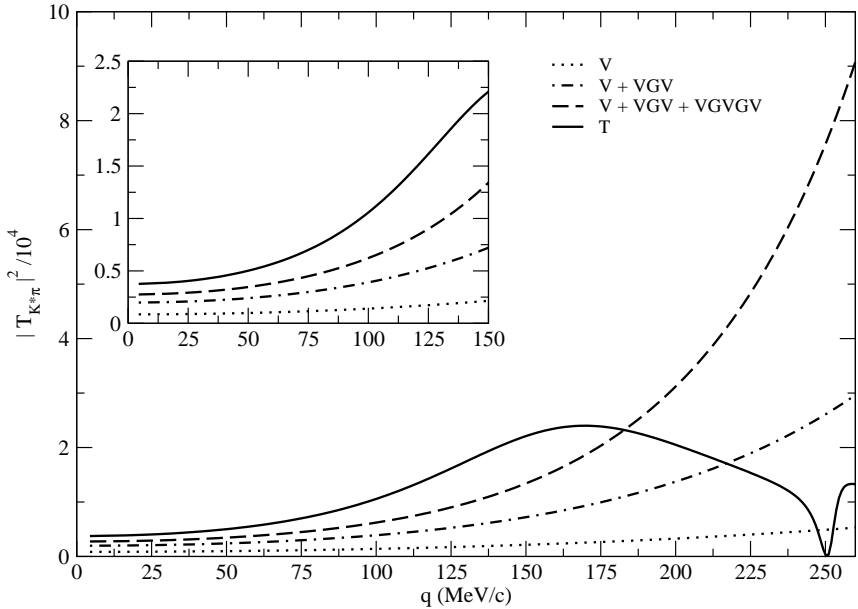}
\caption{Same as Fig.~\ref{fig:perturb1} but 
 with all the allowed $VP$ channels and with finite pseudoscalar
 masses}
\label{fig:perturb2}
\end{center}
\end{figure}

The real case has finite pseudoscalar masses and coupled channels
($\phi K$, $\omega K$, $\rho K$, $K^*\eta$, $K^*\pi$) and one has
different thresholds for different channels. This case is shown
in Fig.~\ref{fig:perturb2}.
 This fact changes
the behavior of the amplitudes close to the $K^*\pi$ threshold,
where they no longer vanish, and, although the perturbative
expansion is seen to converge to the unitary amplitude, the
convergence is now slower. Once more we see a lack of
convergence when we get close to the resonance peak around 
$170\mev/c$.

\begin{figure}[h]
\begin{center}
\includegraphics[width=0.7\textwidth]{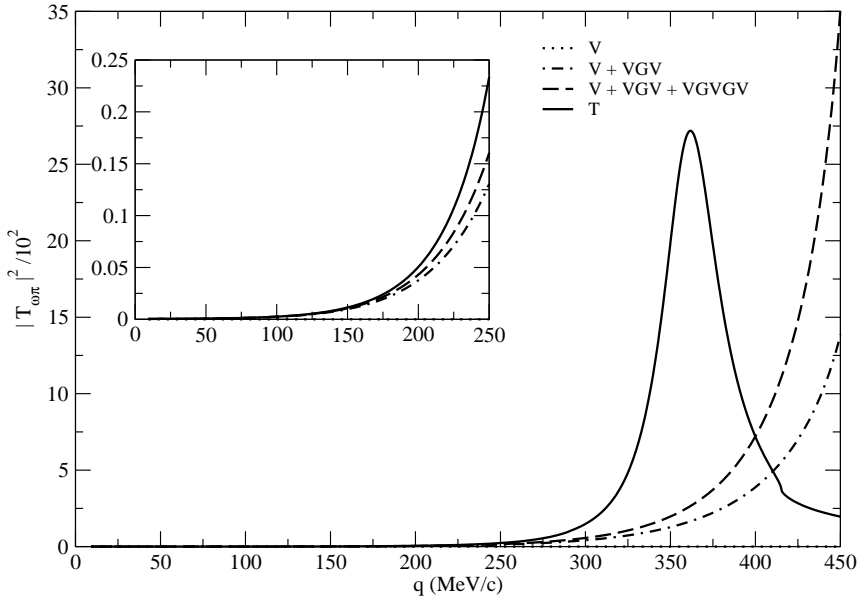}
\caption{Same as Fig.~\ref{fig:perturb2} for the
$\omega\pi\to\omega\pi$ case}
\label{fig:perturb3}
\end{center}
\end{figure}

In Fig.~\ref{fig:perturb3} we show the same amplitudes
 as before for the
$\omega\pi\to\omega\pi$ cases.
 Now, the tree level amplitude is
zero, see Table~\ref{tab:Cij7}, and the first resonance appears
around $350\mev/c$. We see that the convergence of the
perturbative expansion extends up to higher momenta than in the
previous case, up to about $250\mev/c$. A combined conclusion
from these two cases is that the perturbative expansion works
well as far as we are not close to the lowest resonance, and of
course, as far as $q$ is small compared to $\Lambda_\chi$. This
reflects the mathematical theorem of complex variable that a
series converges up to the first singularity, in our case the 
poles in
the complex plane associated to the resonances. It is worth
noting that what limits the convergence of the series expansion
is the appearance of the lowest resonance, which in the present
work appears at momenta much smaller than $\Lambda_\chi$. The
unitary amplitude has no problems of convergence, based as it is
in the $N/D$ method, as far as one used
 a proper interaction and
included all relevant coupled channels. Up to energies of
$\sqrt{s}\simeq 1.5\textrm{ GeV}$ we consider that we are taking
into account the relevant channels. This is about a momentum of
about $500\mev/c$ for th $\rho\pi$ channel, which has the highest
momentum. This momentum is still smaller than the $600\mev/c$
of the $\pi\pi$ interaction, where the chiral unitary approach
was shown to be rather successful in \cite{ollerND}. Nevertheless
we shall comment
in the results sections on possible consequences from the
inclusion of channels other than $VP$. 

Contrary to the
perturbative expansion, the unitary amplitude allows for poles
corresponding to bound states or resonances. In the next section
we address this issue.


\section{Search for poles}

\subsection{Unphysical Riemann sheets}

The association of physical resonances to poles of the
scattering matrix in unphysical Riemann sheets is a very
powerful tool to identify the resonances. The results of the
scattering theory say that bound states reflect as a pole for
$\Ima(q)>0$ and $\Rea(q)=0$, with $q$ the
momentum $q_l$ of Eq.~(\ref{eq:q}), i.e., in the real $s$
axis below the lowest threshold. The resonances can appear only
for $\Ima(q)<0$ which means $s$ with an argument larger than
$2\pi$ and $\Rea(s)$ above the lowest threshold. This is what
we will call second Riemann sheet (R2) for the function $T$ for
the variable $s$. If  these poles are not very far from the
real axis they occur in $\sqrt{s}=(M\pm
i\Gamma/2)\equiv\sqrt{s_p}$ with $M$ and $\Gamma$ the mass and
width of the resonance respectively.  Of course the only
meaningful physical quantity
is the value of the amplitudes for real
$\sqrt{s}$, i.e., the reflexion of the pole on the real axis.
Therefore, only poles not very far from the real axis would be
easily identified experimentally as a resonance.

The effect of passing $s$ to R2 has consequences only for the $G$
functions. To evaluate $G$ in R2 we can use the Schwartz
reflexion theorem which states that if a function $f(z)$ is
analytic in a region of the complex plane including a portion of
the real axis in which $f$ is real, then $[f(z^*)]^*=f(z)$. The
loop function $G_l$ satisfies these conditions, therefore, for 
$\Rea(\sqrt{s})>m_l+M_l$ we have

\be
G_l(\sqrt{s}-i\epsilon)=[G_l(\sqrt{s}+i\epsilon)]^*
=G_l(\sqrt{s}+i\epsilon)-i2\Ima G_l(\sqrt{s}+i\epsilon).
\label{eq:disc1}
\ee

Since the beginning of R2 is equal to the end of R1 we have

\be
G_l^{II}(\sqrt{s}+i\epsilon)=G_l^I(\sqrt{s}-i\epsilon)
=G_l^I(\sqrt{s}+i\epsilon)-i2\Ima G_l^I(\sqrt{s}+i\epsilon).
\label{eq:disc2}
\ee
\noindent
where the superindices $I$ and $II$ refer to R1 and R2 respectively.

The imaginary part of the loop function can be very easily
evaluated from Eq.~(\ref{eq:loop}), for instance with Cutkosky
rules, giving
$\Ima G_l^I(\sqrt{s}+i\epsilon)=-\frac{q}{8\pi\sqrt{s}}$.

In principle Eqs.~(\ref{eq:disc1}) and (\ref{eq:disc2})
are true only very close to the real axis but,
since the analytic continuation to general complex plane is
unique we can write for a general $\sqrt{s}$

\begin{equation}
G_l^{II}(\sqrt{s})=G_l^{I}(\sqrt{s})+i\frac{q}{4\pi\sqrt{s}},
\label{eq:GII}
\end{equation}
\noindent
with $\Ima(q)>0$.
In Eq.~(\ref{eq:GII})  one can use for $G_l^{I}$
 either Eq.~(\ref{propdr}) or the
result of the cutoff method.

One could also have gone to R2 by using Eq.~(\ref{propdr})
but with the solution of $q_l$
with $\Ima(q_l)<0$, but again one finds the problem of the
multivaluedness of the $\ln$ functions. We have checked, by
comparing with the result obtained from Eq.~(\ref{eq:GII}), 
that one can use Eq.~(\ref{propdr}) as it is written with the
prescription of the $\ln$ explained below Eq.~(\ref{propdr}) and
using $\sqrt s$ in the form $a+ib$, $a$ and $b$ real.

When looking for poles we will use $G_l^I(\sqrt{s})$ for
$\Rea(\sqrt{s})<m_l+M_l$ and  $G_l^{II}(\sqrt{s})$ for 
$\Rea(\sqrt{s})>m_l+M_l$.
This prescription gives the pole positions and half widths closer
to those of the corresponding Breit-Wigner forms in the real
axis.
 In this way, when being below the lowest
threshold, we could also obtain possible pure bound states.


\subsection{SU(3) symmetry breaking scan
\label{subsec:SU3scan}}

\begin{figure}[h]
\begin{center}
\includegraphics[width=0.8\textwidth]{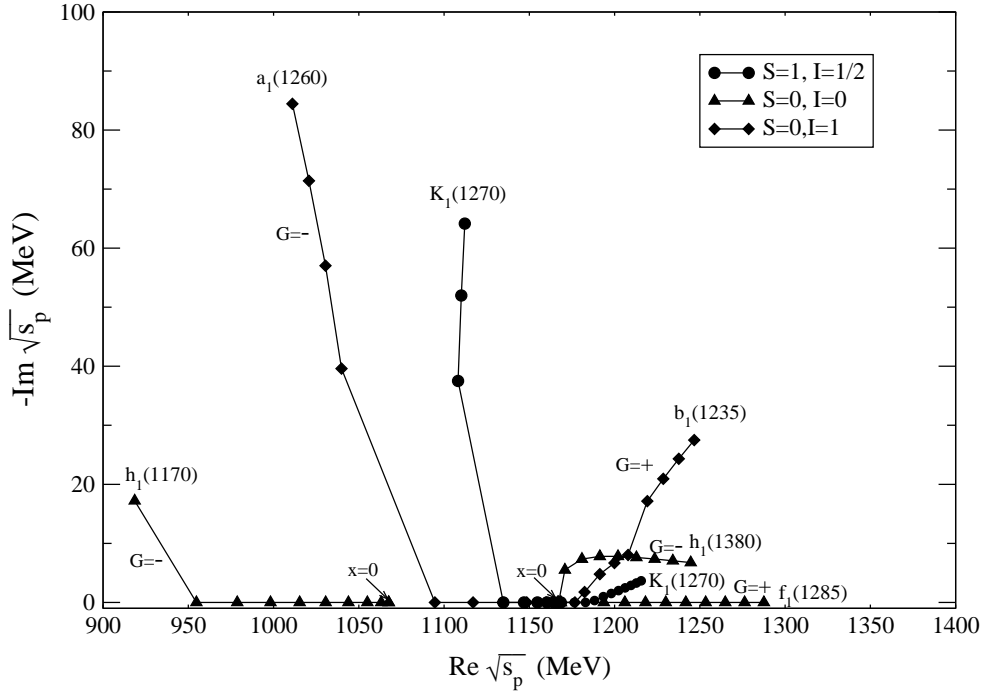}
\caption{Trajectories of the poles in the scattering amplitudes
obtained by increasing the $SU(3)$ breaking parameter, $x$, from
$0$ to $1$. The loop functions are regularized by means of the
subtraction constant method.}
\label{fig:polestraj1}
\end{center}
\end{figure}

The use of physical vector and pseudoscalar masses, both in the
potential $V_{ij}$ as in the loop functions, allows for SU(3)
symmetry breaking. By following a similar procedure to 
Refs.~\cite{sourav,Jido:2003cb}, we are going to look for poles
starting from the exact $SU(3)$ limit and breaking $SU(3)$
symmetry gradually. This can be accomplished by starting using a
unique vector meson mass, $M_0$, and pseudoscalar mass, $m_0$,
(obtained as an average of the physical masses inside each
$SU(3)$ multiplet) and changing the masses in the following way:

\begin{eqnarray}
M_i^2(x)=M_0^2+x(M_i^2-M_0^2) \nonumber \\
m_i^2(x)=m_0^2+x(m_i^2-m_0^2),
\end{eqnarray} 
with $0\le x\le 1$.
In this way, the masses used are the $SU(3)$ averaged masses
for $x=0$ and the physical masses for $x=1$. In
Fig.~\ref{fig:polestraj1} we show the position of the poles for
the $(S,I)$ channels $(0,0)$, $(0,1)$ and $(1,1/2)$, (the only
$(S,I)$ channels for which we find poles),
 evaluating the loop
functions with the subtraction constant method.

In the exact $SU(3)$ limit ($x=0$) we find two degenerate octet
poles in all the $(S,I)$ channels at $\sqrt{s}=(1168+i0)\mev$,
and also a singlet pole for the $(0,0)$ channel at
$\sqrt{s}=(1067+i0)\mev$. Hence, our guess at the end of
subsection~\ref{subsec:treeV} that there could be two octets and
one singlet of dynamically generated resonances gets confirmed.
As we break $SU(3)$ gradually, by increasing $x$, two branches
for each $(S,I)$ channels of the octet and one for the singlet
emerge. The
branches end in the physical mass situation, $x=1$. Note that
each branch for the $(0,0)$ and $(0,1)$ channels has well defined
$G-$parity while this is not the case for $(1,1/2)$ since this
corresponds to non-zero strangeness. In the plot we have also
written for each pole our guess  for the correspondence with
physical $1^+$ resonances of the PDG.
These assignments will be justified in the detailed discussion of 
section~\ref{sec:detail}.

The plot in Fig.~\ref{fig:polestraj1} is instructive but must be
interpreted with caution. In principle, the real part of the pole
position at $x=1$ should reflect the mass of the resonance while
two times the imaginary part should be the width. One may wonder
how stable are the results with respect to reasonable changes in
the regularization scheme. An estimation of the uncertainties can
be done by using the regularization method of fixing $G=0$ at the
vector mass \cite{lutzaxials}. The results are qualitatively
similar for the trajectories and the real part for the pole
positions at $x=1$ differ in the worst of the cases in less than
$100\mev$ and in most cases by less than $30\mev$. The imaginary
part of the pole positions also differ in similar amounts, but
relative to the absolute values of the mass and width of the
resonances, the differences in the width are more significant.
Yet, here we must observe that the width obtained so far is only a
first approximation for the following reasons:

a) We have not considered other decay channels which are not made
of a vector and a pseudoscalar. Other channels to which the
resonance couples weakly can in practice give a sizeable partial
decay width because of the large phase space available. For
example, this would be the case of the $b_1(1235)$ going to four
pions, where the phase space is favored with respect to our $VP$
channels. While including these extra channels in the coupled
channels approach goes beyond the scope of the present work, it
is important to note that the weaker strength of the couplings to
these channels makes their repercussion on the real part of the
pole positions less relevant since there are no restrictions of
phase space for the real part of the amplitudes. 
Hence, we might expect small changes in he real part of the pole
positions from these neglected sources, however, not altering the
important fact that these poles appear for these quantum numbers.

b) The second reason is that so far we have not considered the
width of the vector mesons in their propagators in the loop
functions. For the case of the $\rho$ and the $K^*$ it is important
to take this into consideration. We shall take this into account
in the results section and we shall see that this modifies the
width of the resonances but only very slightly their mass. We will
also use another method to account for the width of the vector
mesons by using the coupling of the resonances to the $VP$
channels and evaluating the partial decay widths by means of a 
convolution with the mass distribution of the vector and
axial-vector mesons. This will be seen in the next chapter.


\subsection{Couplings and partial decay widths}

\begin{figure}[h]
\begin{center}
\includegraphics[width=0.3\textwidth]{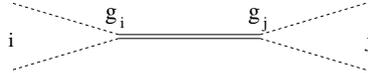}
\caption{Effective interpretation of the $PV\to PV$ scattering
process close to a pole.}
\label{fig:effT}
\end{center}
\end{figure}

The physical interpretation of the poles comes clearer if we
realize that close to a pole, and if it is not very close to a
threshold, the amplitude takes the form of a Breit-Wigner
structure. By looking at the structure of the covariant amplitude
in the Appendix~II Eq.~(\ref{eq:ab})  and the fact that the poles
appear in the $Vb(1-b)$ term, close to a pole the amplitude has
the form (in one channel for simplicity) 
\be
T\simeq\frac{Vb}{1-b}\left(g^{\mu\nu}-\frac{P^\mu P^\nu}{P^2}\right)
\epsilon_\mu\epsilon'_\nu, 
\ee 
\noindent 
which has the structure of a resonant pole amplitude (see
Fig.~\ref{fig:effT})
\be
T\simeq g^2\frac{1}{P^2-M_R^2}
\left(-g^{\mu\nu}+\frac{P^\mu P^\nu}{M_R^2}\right)
\epsilon_\mu\epsilon'_\nu.
\ee 
\noindent 
Generalizing to different channels and considering only the 
transverse
polarizations, the amplitudes close to a pole in the second
Riemann sheet can be expressed as 

\begin{equation}
T_{ij}\simeq \frac{g_ig_j}{s-s_p},
\end{equation}
\noindent
where we have omitted the trivial
$\vec\epsilon\cdot\vec\epsilon\,'$ factor.
Hence the factors $g_i$, which stand as the effective coupling of
the dynamically generated axial-vector resonances to the channel
$i$, can be calculated from the residues of the amplitudes at the
complex poles.

With the values of the couplings we can evaluate the partial
decay widths of the axial-vector mesons into each $VP$
channels, which reads

\begin{equation}
\Gamma_{A\to VP}=\frac{|g_{VP}|^2}{8\pi M_A^2}q.
\label{eq:GAVP}
\end{equation}

In the case where there is little phase space
for the decay or it takes place due to the width of the
particles, we fold the expression
for the width with the mass distribution of the particles
as 

\begin{eqnarray} \nonumber
\Gamma_{A\to VP}&=&\frac{1}{\pi^2} 
\int_{(M_A-2\Gamma_A)^2}^{(M_A+2\Gamma_A)^2} ds_A
\int_{(M_V-2\Gamma_V)^2}^{(M_V+2\Gamma_V)^2} ds_V 
\,\,\Gamma_{AVP}(\sqrt{s_A},\sqrt{s_V}) \\ \nonumber
&\cdot&Im \left\{\frac{1}{s_A-M_A^2+iM_A\Gamma_A}\right\}
Im \left\{\frac{1}{s_V-M_V^2+iM_V\Gamma_V}\right\} \\
&\cdot&\Theta(\sqrt{s_A}-\sqrt{s_V}-M_P)
\Theta(\sqrt{s_A}-\sqrt{s_A^{th}})
\Theta(\sqrt{s_V}-\sqrt{s_V^{th}}),
\label{eq:convolution}
\end{eqnarray}
 
\noindent
where $\Theta$ is the step function, 
$\Gamma_{AVP}=\frac{|g_{VP}|^2}{8\pi s_A}q$ with
$q=\frac{1}{2\sqrt{s_A}}\lambda^{1/2}(s_A,s_V,M_P^2)$, $\Gamma_A$
and $\Gamma_V$ are the axial and vector mesons total width and
$\sqrt{s_A^{th}}$, $\sqrt{s_V^{th}}$, are the threshold energies for the
dominant $A$ and $V$ decay channels. 

In Eq.~(\ref{eq:convolution}) the convolution is done using the
masses and total widths of the particles taken from the PDG.


\section{Detailed study of the $1^+$ resonances
\label{sec:detail}}

\subsection{$S=0$, $I=0$}

\begin{figure}[h]
\begin{center}
\includegraphics[width=1.\textwidth]{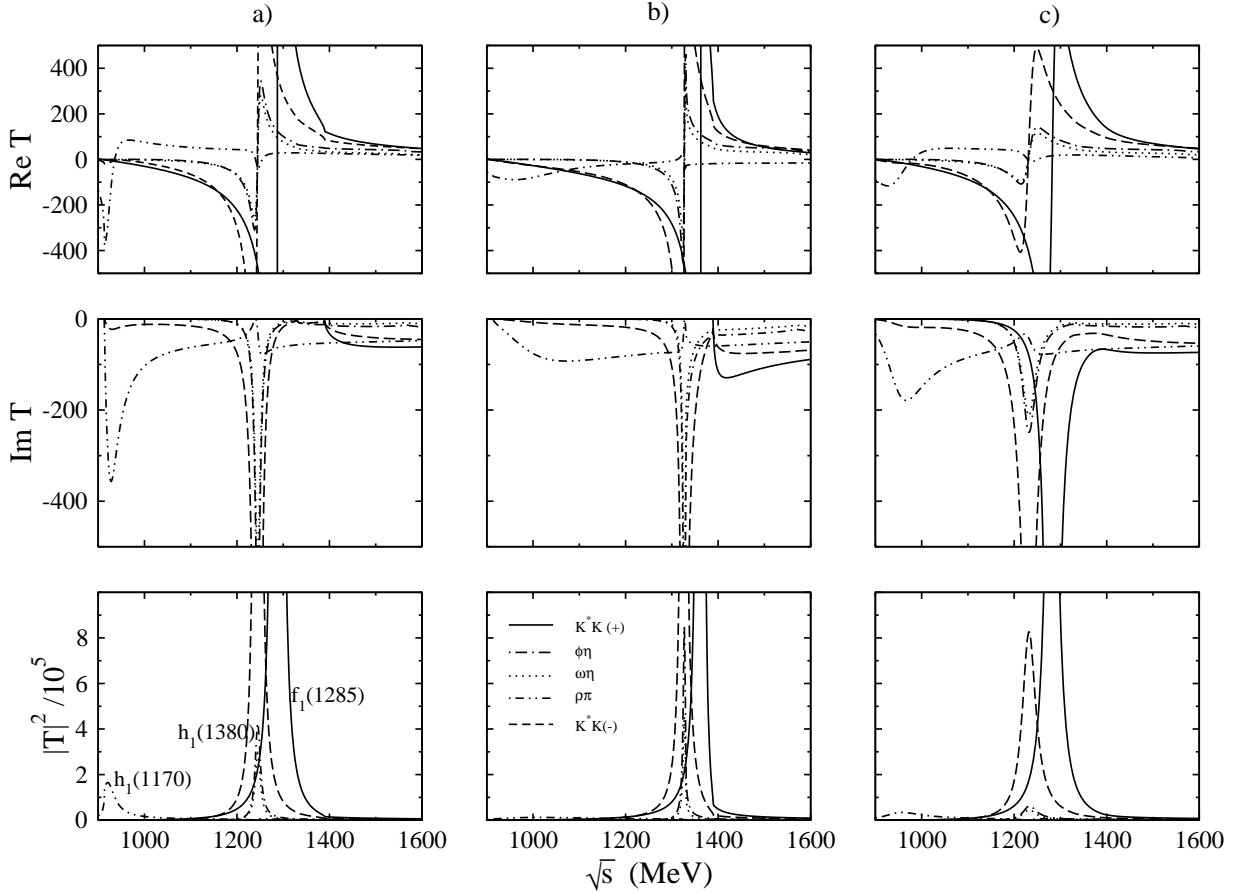}
\caption{Amplitude for $S=0$, $I=0$, for 
the different allowed channels. Left column:
subtraction constant method in the loop functions;
 central column: with the prescription of
$G=0$ at the vector meson mass energy;
right column: cutoff method considering
vector meson width in the propagator.}
\label{fig:T6}
\end{center}
\end{figure}

In Fig.~\ref{fig:T6} we show the results for the diagonal
$T_{ii}$ matrix elements as a function of the energy with the
loop functions evaluated with the subtraction constant method (we
will call it $a$ method) (left column), 
with the prescription of
$G=0$ at the vector meson mass energy ($b$ method)
(central column) and with the cutoff method including the widths
of the vector mesons in the loops ($c$ method) (right column). 
This latter method evaluates Eq.~(\ref{eq:loop}) performing the
$q^0$ integration analytically and the $\vec{q}$ integration with a
cutoff $q_{max}=1000\mev$ substituting $M_l^2\to
M_l^2-iM_l\Gamma$.

We do not show the results with the cutoff method
without widths in the vector meson propagator in the loops since
they are very similar to the ones with the $a$
method (actually, the subtraction constant has been chosen to 
agree with the cutoff method).
We can clearly see the reflexion on the real axis
of the poles found in Fig.~\ref{fig:polestraj1} in this $(0,0)$ channel,
which corresponds to the $a$ method.
The $a$ and $b$ methods give qualitative similar results but there
are significant differences at low energies although not so
large at higher energies. 
When we include the width in the vector meson propagators of the
loops, we obtain a broadening of the amplitudes.
All these discrepancies give an idea of the
uncertainty we have in the model.

In Table~\ref{tab:poles6} we show the position of the poles
obtained with the $a$ method  and the couplings to the different
isospin channels. 

\begin{table}[h]
\begin{center}
\begin{tabular}{|c|cc|cc|cc|}
\hline
 $\sqrt{s_p}$ & \multicolumn{2}{c|}{$919-i17$} &
\multicolumn{2}{c|}{ $1245-i7$} & \multicolumn{2}{c|}{ $1288-i0$}\\
\cline{2-7} &&&&&& \\[-4.5mm]
& $g_i$ & $|g_i|$ & $g_i$ & $|g_i|$& $g_i$ & $|g_i|$ \\
\hline &&&&&& \\[-4.5mm]
$\frac{1}{\sqrt{2}}(\bar K^*K+K^*\bar K)$   & $-$ & $-$ & $-$ & $-$& $7230+i0$ & $7230$ \\
$\phi\eta$ & $46-i13$& $48$ & $-3311+i47$& $3311$  &   $-$ & $-$\\
$\omega\eta$ & $23-i28$& $36$ & $3020-i22$& $3020$  &  $-$ & $-$\\ 
$\rho\pi$ & $-3453+i1681$ & $3840$& $648-i959$& $1157$ & $-$ & $-$\\
$\frac{1}{\sqrt{2}}(\bar K^*K-K^*\bar K)$  & $781-i498$ & $926$ & $6147+i183$ & $6150$ & $-$  & $-$\\
\hline
\end{tabular}
\caption{Pole positions and couplings for $S=0$, $I=0$. (All the
units are $\mev$.)}
\label{tab:poles6}
\end{center}
\end{table}

The pole at $919-i17\mev$, which is found for the channels with 
negative $G-$parity, may be identified with the 
$h_1(1170)$  resonance. In
the PDG the only decay channel seen so far is $ \rho \pi$,
although we find this pole in all the channels allowed by the
quantum numbers, even if they are kinematically closed.
However, in Fig.~\ref{fig:T6} these other channels 
are difficult to see since the
amplitude is dominated in this region by the $\rho\pi$ channel.

The partial widths obtained for the 
axial-vector resonance decaying into $VP$
are presented in Table~\ref{tab:decays6}.
\begin{table}[h]
\begin{center}
\begin{tabular}{|c|c|c|c|c|}
\hline &&& \\[-4.5mm]
& $\Gamma_{tot}^{exp}$ & $\Gamma_i^{exp}$ & $\Gamma_i^{th}(a)$ & $\Gamma_i^{th}(b)$\\
\hline &&& \\[-4.5mm]
$h_1(1170)\to \bar K^*K+c.c.$   & $360\pm40$ & $-$    & $0.5$ & $1.4$\\
$h_1(1170)\to \rho\pi$          & $''$       & $seen$ & $77$  & $115$ \\
$h_1(1170)\to \omega\eta$       & $''$       & $-$    & $0$   & $0$  \\ 
$h_1(1170)\to \phi\eta$         & $''$       & $-$    & $0$   & $0$  \\
\hline &&& \\[-4.5mm]
$h_1(1380)\to \bar K^*K+c.c.$   & $91\pm30$  & $seen$ & $45$  & $36$\\
$h_1(1380)\to \rho\pi$          & $''$       & $-$    & $9$   & $4$ \\
$h_1(1380)\to \omega\eta$       & $''$       & $-$    & $25$  & $16$\\ 
$h_1(1380)\to \phi\eta$         & $''$       & $-$    & $0$   & $0$ \\
\hline &&& \\[-4.5mm]
$f_1(1285)\to \bar K^*K+c.c.$   & $24\pm1$   & $not\,seen$ & $0.3$ & $0.2$ \\
\hline
\end{tabular}
\caption{Partial decay widths for $S=0$, $I=0$. (All the
units are $\mev$.)}
\label{tab:decays6}
\end{center}
\end{table}
\noindent
The results that we obtain are compatible with the scarce
experimental information in the sense that the only decay
channel seen  is the $\rho\pi$, which is the only relevant in
our case. We can see that the widths for this  channel
calculated with the $a$ and $b$ methods are qualitatively
similar.
It is also instructive to see that with 
the $c$ method (see 3rd column
of Fig.~\ref{fig:T6}) we see the apparent full width  of the
resonance (around $90\mev$), which is in reasonable agreement
with the sum of all the partial decay widths in 
Table~\ref{tab:decays6} for the $h_1(1170)$.

The $1245-i7\mev$ pole can be assigned to the $h_1(1380)$
resonance.
Note that in the PDG the isospin of this resonance is not given,
although in the classification schemes it is assumed to have
$I=0$. Our assignment, as well as suggested in \cite{lutzaxials},
is also $I=0$.

The only experimentally observed decay channel of the 
$h_1(1380)$ is  $\bar K^*K$ (with $G-$parity negative). In
Table~\ref{tab:decays6} we see our results for the different 
partial decay widths. It is difficult to extract conclusions
from Table~\ref{tab:decays6} given the scarce experimental 
data, but we should stress that the only channel experimentally
seen is precisely the dominant one in our calculations. In this
case the sum of our partial $VP$ decay channels is compatible
with the total experimental width. 
We also see that method $c$ gives a width (around $40\mev$)
qualitatively similar to the sum of the partial decay widths in 
Table~\ref{tab:decays6}.

The pole at $1288-i0\mev$ is below the $\bar K^*K$ threshold
which is the only allowed $PV$ channel for positive $G-$parity,
and therefore the pole appears as a bound state of $\bar K^* K +
c.c.$. Actually, in Fig.~\ref{fig:T6} with the $a$ and $b$
methods $|T|^2$ goes to infinity at the pole position. It is
reasonable to  assign this pole to the $f_1(1285)$ resonance.
The reason why we do not get width for this resonance, while in
the PDG it is said to be $24\mev$, is because there are other
decay channels different to $VP$ that we are not obviously
considering, (like  $4\pi$ ($33$\%), $\eta\pi\pi$ ($52$\%) or
$K\bar K\pi$ ($10$\%)) . In fact the  $\bar K^* K + c.c.$ is
quoted in the PDG as "not seen" which agrees, in a first
approximation, with the result obtained here as a bound state.
If we evaluate the partial decay width  into this channel, with
the coupling obtained and with the convolution of the widths we
get about $0.3\mev$ which justifies why this decay channel has
so far not been seen.
 This small number is actually rather
unstable since it can be made even an order of magnitude bigger
by enlarging the limits in the convolution of
Eq.~(\ref{eq:convolution}). 
The apparent width from the $|T|^2$
plot when considering the vector meson width in the propagator
of the loop functions is around $20\mev$. 
With all these apparent instabilities in the widths, one should
not lose the important point which is that, comparatively to the
other resonances, the width for this resonance is very small.
 It is
worth noting that the $f_1(1285)$ couples only to $K^*\bar K$ in
our theory and with such a large strength qualifies strongly as
a quasibound $K^*\bar K$ or $\bar K^*K$ state. The small
experimental total width indicates a small coupling to other
channels which are kinematically open and barely change the
nature of the resonance as a bound system of $K^*\bar K$.

When arriving to this point it is important to note that there
is experimentally another resonance, the $f_1(1420)$, which is
assigned by the PDG to the $1^{++}$ nonet. However,
in our scheme this resonance has no counterpart. This is
because, as explained in subsections~\ref{subsec:treeV} and
\ref{subsec:SU3scan}, the interaction of two octets give one
singlet plus two octets (three poles), therefore there is no
way, with the interaction of one vector and one pseudoscalar
mesons, to generate dynamically one more  pole in this channel
and, therefore, there is no room for one more resonance like the
$f_1(1420)$. Therefore, in our scheme, the $f_1(1420)$ cannot be
considered as a dynamically generated resonance from the $PV$
interaction. 

On the other hand, the $h_1(1170)$ state that we generate (the
original singlet in Fig.~\ref{fig:polestraj1}) has the same quantum
numbers as the  isoscalar member of the octet of the $b_1(1235)$,
the $h_1(1380)$, and hence the set of the $b_1(1235)$, $h_1(1170)$
and $h_1(1380)$, together with the strange partners that we will
discuss below, agree with the association of a nonet to all these
states of the PDG.


\subsection{$S=0$, $I=1$}

\begin{figure}[h]
\begin{center}
\includegraphics[width=1.\textwidth]{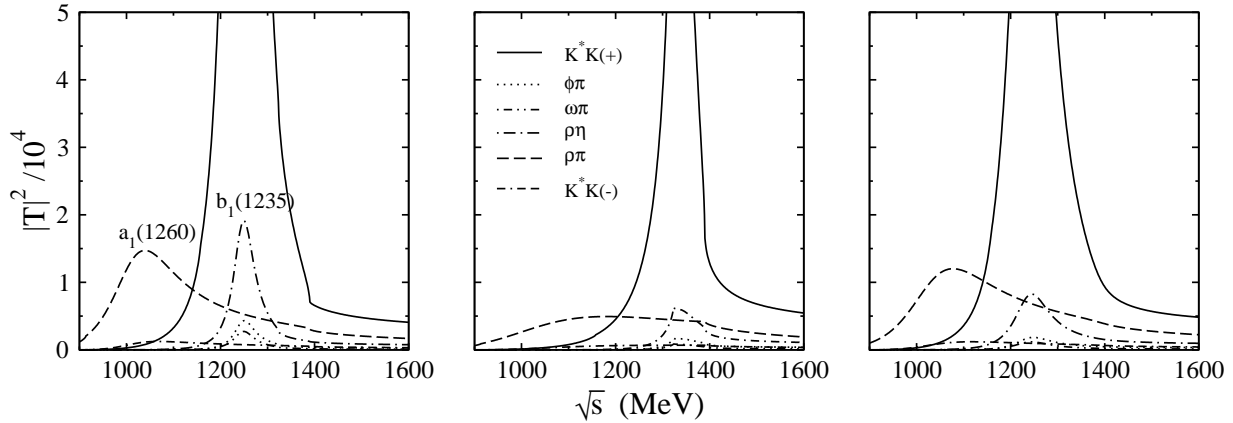}
\caption{Amplitude for $S=0$, $I=1$, for the different channels.
Left plot:
subtraction constant method in the loop functions;
  central plot: with the prescription of
$G=0$ at the vector meson mass energy; right plot: cutoff method considering
vector meson width in the propagator.}
\label{fig:T7}
\end{center}
\end{figure}

In Fig.~\ref{fig:T7} we show the results for the $|T|^2$ with 
the same three methods used in the previous subsection.

 We clearly see the reflexions in the real axis of the poles found
in the complex plane, (shown in Table~\ref{tab:poles7}). 

\begin{table}[h]
\begin{center}
\begin{tabular}{|c|cc|cc|}
\hline
 $\sqrt{s_p}$ & \multicolumn{2}{c|}{$1011-i84$} &
\multicolumn{2}{c|}{$1247-i28$ } \\
\cline{2-5} &&&& \\[-4.5mm]
& $g_i$ & $|g_i|$ & $g_i$ & $|g_i|$ \\
\hline &&&& \\[-4.5mm]
$\frac{1}{\sqrt{2}}(\bar K^*K+K^*\bar K)$    & $-$& $-$ & $6172-i75$ & $6172$  \\
$\phi\pi$   & $-$& $-$ & $2087-i385$   & $2122$\\
$\omega\pi$ & $-$& $-$ & $-1869+i300$  & $1893$\\ 
$\rho\eta$  & $-$& $-$ & $-3041+i498$  & $3082$\\
$\rho\pi$   & $-3795+i2330$& $4453$ &  $-$  & $-$\\
$\frac{1}{\sqrt{2}}(\bar K^*K-K^*\bar K)$  & $1872-i1486$&$2390$ & $-$& $-$ \\
\hline 
\end{tabular}
\caption{Pole positions and couplings for $S=0$, $I=1$.}
\label{tab:poles7}
\end{center}
\end{table}

The pole at $1247-i28\mev$, which is found for the channels with 
positive $G-$parity,
 may  be assigned to the $b_1(1235)$
resonance. In the PDG the full width is quoted to be $142\mev$ and
the decay channels are quoted as $\omega\pi$ (dominant),
$\rho\eta$ (seen), $4\pi$ ($<50$\%), $K\bar K\pi$ ($<14$\%)
and $\phi\pi$ ($<1.5$\%), (See Table~\ref{tab:decays7}).
\begin{table}[h]
\begin{center}
\begin{tabular}{|c|c|c|c|c|}
\hline &&& \\[-4.5mm]
& $\Gamma_{tot}^{exp}$ & $\Gamma_i^{exp}$ & $\Gamma_i^{th}(a)$ & $\Gamma_i^{th}(a)$\\
\hline &&& \\[-4.5mm]
$b_1(1235)\to \bar K^*K+c.c.$  & $142\pm9$  & $-$        & $7$  & $10$  \\
$b_1(1235)\to \phi\pi$         & $''$       & $<1.5$\%   & $12$ & $13$  \\
$b_1(1235)\to \omega\pi$       & $''$       & $dominant$ & $25$ & $25$  \\ 
$b_1(1235)\to \rho\eta$        & $''$       & $seen$     & $8$  & $9$   \\
\hline &&& \\[-4.5mm]
$a_1(1260)\to \rho\pi$         & $250-600$  & $seen(60$\%$?)$  & $106$ & $156$ \\
$a_1(1260)\to \bar K^*K+c.c.$  & $''$       & $seen(<10$\%$?)$ & $6$   & $11$  \\
\hline
\end{tabular}
\caption{Partial decay widths for $S=0$, $I=1$. (All the
units are $\mev$.)}
\label{tab:decays7}
\end{center}
\end{table} 
\noindent
Therefore one could expect that the $VP$ decay channels 
would account for around $50$\% of the total width. This is
compatible with what we 
obtain, as can be seen in Table~\ref{tab:decays7}.
Note also that the channel quoted as dominant in the PDG is
indeed the dominant one in our results. It is also interesting to
call the attention to the fact that the strongest coupling of the
$b_1(1235)$ is to $K^*\bar K$ as it was the case of the
$h_1(1380)$, consistently with the fact that they belong to the
same $SU(3)$ octet in the $SU(3)$ limit. It is interesting to see
that this feature remains in spite of the $SU(3)$ breaking.
The method $c$ provides an apparent total width of about
$80\mev$.

The pole at $1011-i84\mev$, which is found for the channels with 
negative $G-$parity, could be assigned to the $a_1(1260)$
resonance.
 Note that the values for the mass and width quoted for
this resonance in the PDG, $1230\pm40$ and $250-600\mev$
respectively,  have a large uncertainty. This can justify the
discrepancy in the position of the pole we obtain. For the
partial decays widths there is no good data quoted in the PDG
(every one of the many decay channels are just quoted 
as "seen"). However, in the detailed explanation in the PDG, some
experiment gives for $\rho\pi$  $60$\% and for $\bar K^*K+c.c.$
$8-15\mev$\%, what would be in reasonable agreement
with what we obtain.
The method $c$ provides an apparent width of about $300\mev$
which is tied to he slow fall down of $|T|^2$ at the right of the
peak, which differ from a typical Breit-Wigner shape.

It is interesting to note that the octet of  
the $a_1(1260)$ resonance has 
been considered as one of the fundamental fields in chiral
theories which deal explicitly with spin-1 fields \cite{egpdr}.
In this latter work
 the $L_i$ parameters of \cite{gassleut} have been derived
assuming exchange of these vector mesons (and scalars). The
comment is worth making because the use of the chiral interaction
between the pseudoscalar and vector mesons used here, together
with the unitarization procedure, generates dynamically
 the $a_1(1260)$
resonance and, hence,
 introducing it further as an explicit degree of
freedom would lead to doublecounting.

There is another observation worth making which is the fact that
there is another $a_1$ resonance in the PDG at higher energies,
the $a_1(1640)$, which we do not generate dynamically and is
more likely to be a genuine $q\bar{q}$ meson. At the same time
there is another observation to make which is the fact that no
other $b_1$ resonance is quoted in the PDG.

\subsection{$S=1$, $I=1/2$}

\begin{figure}[h]
\begin{center}
\includegraphics[width=1.\textwidth]{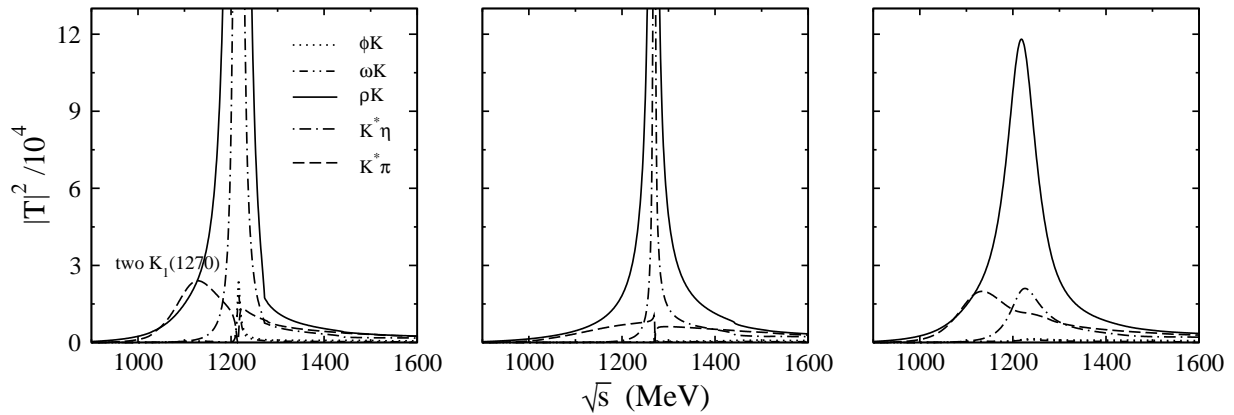}
\caption{Amplitude for $S=1$, $I=1/2$, for the different channels.
Left plot:
subtraction constant method in the loop functions;
  central plot: with the prescription of
$G=0$ at the vector meson mass energy; right plot: cutoff method considering
vector meson width in the propagator.}
\label{fig:T5}
\end{center}
\end{figure}

In Fig.~\ref{fig:T5} we show the results for the amplitudes with
the same three methods considered in the other channels.

In Table~\ref{tab:poles5} we show the pole positions and the
couplings obtained.
\begin{table}[h]
\begin{center}
\begin{tabular}{|c|cc|cc|}
\hline &&&& \\[-4.5mm]
 $\sqrt{s_p}$ & \multicolumn{2}{c|}{$1112-i64$} &\multicolumn{2}{c|}{ $1216-i4$ } \\
\cline{2-5}
& $g_i$ & $|g_i|$ & $g_i$ & $|g_i|$ \\
\hline  &&&& \\[-4.5mm]
$\phi K$   &  $1587-i872$  & $1811$ & $1097-i400$  & $1168$  \\
$\omega K$ &  $-1860+i649$ & $1970$ & $-1033+i375$ & $1099$  \\
$\rho K$   &  $-1524+i1154$& $1912$ & $5274+i297$  & $5282$  \\ 
$K^* \eta$ &  $27+i155$    & $157$  & $3459-i95$   & $3460$  \\
$K^* \pi$  &  $4187-i2098$ & $4683$ & $340-i984$   & $1041$  \\
\hline
\end{tabular}
\caption{Pole positions and couplings for $S=1$, $I=1/2$.}
\label{tab:poles5}
\end{center}
\end{table}
The lowest pole couples strongly to $K^*\pi$, very weakly to
$K^*\eta$ and moderately to the rest of channels.
On the contrary, the higher pole couples strongly to $\rho K$ and
$K^*\eta$ while moderately to the other channels.

\begin{figure}[h]
\begin{center}

\includegraphics[width=1\textwidth]{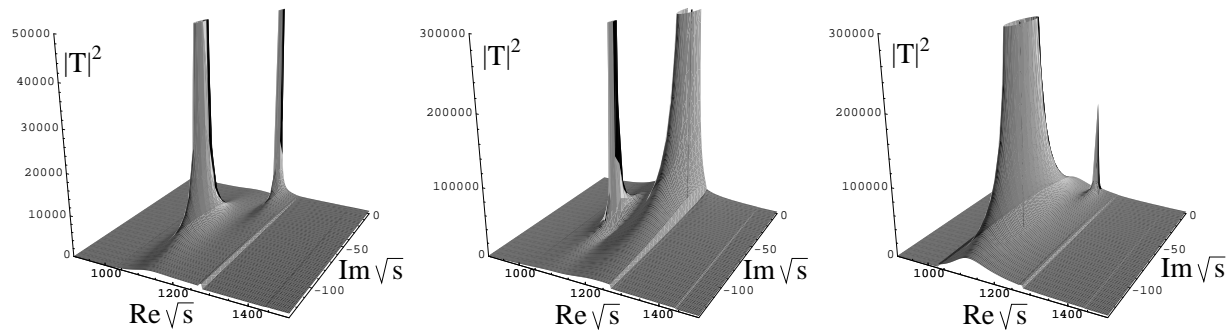}
\caption{$|T|^2$ in the second Riemann
sheet for $\phi K$, $\rho K$
and $K^*\pi$ channels for $S=1$, $I=1/2$. }
\label{fig:poles3D}
\end{center}
\end{figure}

In Fig.~\ref{fig:poles3D} we plot, as an illustrative example,
the modulus squared of the scattering matrix in the second
Riemann sheet for three different channels. The pole structure
can be clearly seen as well as the relative strength of each
channel in the two different poles (similar plots can be obtained
for the rest of channels).

In the PDG there are two physical $S=1$, $I=1/2$, resonances with
$J^{P}=1^+$, which are the $K_1(1270)$ and the $K_1(1400)$.
Actually these two resonances have been usually considered to be
a mixture of the $K_1$ members of the $1^{++}$ and $1^{+-}$
octets, called $K_{1A}$ and $K_{1B}$ respectively.
At this point we have difficulty assigning the poles found to
these resonances. In order to have a feeling of which reasonable
assignment to make we study the partial decay widths of these two
resonances. This is shown in Table~\ref{tab:decays5}. In the
table we have assumed for convolution purposes and phase space
the mass of the resonance to be $1270\mev$ and in the last column
we show the expected partial decay width coming from the poles in 
Table~\ref{tab:poles5}. We can see that the resonance at
$1112\mev$ couples strongly to the $K^*\pi$ and leads to a partial
decay width in this channel of $113\mev$ (around $150\mev$ should
we considered the mass to be $1400\mev$). On the other hand, the
$1216\mev$ resonance couples dominantly to  $\rho K$. In view of
this, one would be tempted to assign the $1112\mev$ pole to the
$K_1(1400)$ resonance and the $1216\mev$ pole to the $K_1(1270)$
since this is the experimental case. Yet, the difference of
$300\mev$ in the mass in the case of the $K_1(1400)$ is not an
appealing feature, since  this difference is much larger than the
$K_1(1400)$  width ($174\mev$). There is another possible
scenario which we find more appealing. The detailed explanation
of the PDG on the determination of the decay widths shows a clear
discrepancy between different methods of determination of the
$K_1(1270)$ width. In particular, a set of experiments using $K$
beams leads to much larger widths (about a factor three) than
another set that uses pion beams. This could find an explanation
if one assumed that the experimental $K_1(1270)$ resonance is a
superposition of two resonances that couple with different
strength to a particular channel. This is indeed the case for
the two poles that we find, one of them coupling strongly to
$K^*\pi$ and the other one to $\rho K$. Different experiments
which favor mechanisms that give a bigger weight to one of these
channels would lead to very different widths of the resonance.
This is the case with the recent findings
 that the $\Lambda(1405)$
resonance corresponds to actually two poles and different
experiments favor one or the other pole, leading to different
visible widths of the resonance \cite{Jido:2003cb,nieves,hyodo}.

There are some experimental features in the different
experiments that could be understood within the two pole
structure that we obtain. Indeed, some experiments provide a
dominance of the $K_1(1270)$ decay into $\rho K$ 
\cite{gavillet,rodeback,crennell} while other experiments show a
clear dominance of the $K^*\pi$ decay mode \cite{firestone}. In
our theoretical approach the dominance of the $\rho K$ decay in
a reaction has to be interpreted as a sign that the dynamics of
the reaction favours the second pole (which couples mostly to
$\rho K$) and this would have a small width. This is indeed the
case experimentally, and the total width is about $60\mev$
(similar to our results). On the other hand, the dominance of
the $K^*\pi$ decay channel in another experiment would have to
be interpreted as the dynamics of this reaction favouring the
coupling to the first resonance. In this case the total width
would be large and indeed this is what is seen in the experiment
of \cite{firestone} where the total width is about $190\mev$ (we
obtain $125\mev$ with the $a$ method or $200\mev$ with the $b$
method).

In Ref.~\cite{lutzaxials} a broad bump in the speed plot was
associated to the $K_1(1400)$ resonance but we find no pole in
that amplitude in that region.

\begin{table}[h]
\begin{center}
\begin{tabular}{|c|c|c|c|c|}
\hline &&& \\[-4.5mm]
& $\Gamma_{tot}^{exp}$ & $\Gamma_i^{exp}$ & $\Gamma_i^{th}(a)$  & $\Gamma_i^{th}(b)$ \\
\hline &&& \\[-4.5mm]
$K_1(1270)\to \phi K$   & $90\pm20$ & $-$    & $(0,0)$   & $(0,0)$  \\
$K_1(1270)\to \omega K$ & $''$      & $11$\% & $(5,2)$   & $(10,0.1)$   \\
$K_1(1270)\to \rho K$   & $''$      & $42$\% & $(7,52)$  & $(5,12)$  \\ 
$K_1(1270)\to K^*\eta$  & $''$      & $-$    & $(0,0)$   & $(0,0)$  \\
$K_1(1270)\to K^*\pi$   & $''$      & $16$\% & $(113,6)$ & $(194,1)$  \\
\hline
\end{tabular}
\caption{Partial decay widths for $S=1$, $I=1/2$. 
In the last column, the first element is the value obtained
with the coupling to the lower pole
and the second element is the result with the coupling to the 
higher pole. (All the
units are $\mev$.)}
\label{tab:decays5}
\end{center}
\end{table} 
\noindent

It is interesting to see that, when breaking the $SU(3)$
symmetry, the $S=0$ states of the two octets  in
Fig.~\ref{fig:polestraj1} do not mix because they have well
defined $G-$parity (this is opposite to the case of the meson
baryon interaction \cite{Jido:2003cb} where one did not have this
constraint). However, when we go to $S\ne 0$, then the states
are no longer eigenstates of $G-$parity and the octets mix.
This is a well known case for the $K_1$ axial-vectors where
there is much discussion about the mixing in the
literature\footnote{For a review of the theoretical status on
this mixing angle see, for instance, the introduction of
Ref.~\cite{axials}.}.

We next proceed to see which is the mixing angle for the two
poles in $S=1$, $I=1/2$, that we find. We write

\ba \nonumber
|K_1(1270),1>&=&\sin\theta |K_{1A}>+\cos\theta |K_{1B}>,\\
|K_1(1270),2>&=&\cos\theta |K_{1A}>-\sin\theta |K_{1B}>,
\label{eq:mixang}
\ea
\noindent
where $|K_1(1270),1>$ is the state associated to the lower mass
pole and  $|K_1(1270),2>$ to the upper mass pole. In 
Eq.~(\ref{eq:mixang})
$|K_{1A}>$  corresponds to the $SU(3)$
$|8_a,1,1/2,+1/2>$ state and
$|K_{1B}>$ to the $|8_s,1,1/2,+1/2>$ state, 
(in the notation
$|irrep,Y,I,I_3>$, with $irrep$ being the name of the
irreducible representation of $SU(3)$, $Y$ the hypercharge, $I$
the total isospin and $I_3$ the third component of $I$). 
(Note that we have
chosen $I_3=+1/2$ although the following discussion works
obviously for any allowed $I_3$).

\begin{figure}[h]
\begin{center}
\includegraphics[width=0.2\textwidth]{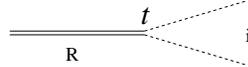}
\end{center}
\caption{Schematic interpretation of the coupling of the
dynamically generated axial-vector resonance to a $PV$ state.}
\label{fig:figcoup}
\end{figure}

The couplings considered so far are (see
Fig.~\ref{fig:figcoup}), by definition,
\begin{equation}
g_i\equiv<R|t|i (I=1/2,I_3=+1/2)>,
\end{equation}
\noindent
where $|R>$ is the generic name for the physical resonance
associated to either pole.
By making use of $SU(3)$ Clebsch-Gordan coefficients, we can
obtain the relation of the couplings in isospin base to 
the ones in $SU(3)$ base, giving

\begin{equation}
\left(\begin{array}{c}
\vspace{0.13cm} g_{\phi K}\\
 \vspace{0.13cm} g_{\rho K}\\
 \vspace{0.13cm}g_{K^*\eta}\\
 \vspace{0.13cm}g_{K^*\pi}\\
\end{array}
\right)
=\left(\begin{array}{cccc}
\vspace{0.1cm}
 -\sqrt{\frac{3}{10}} & \frac{1}{\sqrt{6}} & \frac{1}{\sqrt{30}}  & \frac{1}{\sqrt{6}}\\
 \vspace{0.1cm}
 -\frac{1}{2\sqrt{5}} & -\frac{1}{2}       & -\frac{3}{2\sqrt{5}} & \frac{1}{2}       \\
 \vspace{0.1cm}
 \frac{3}{2\sqrt{5}}  & \frac{1}{2}        & -\frac{1}{2\sqrt{5}} & \frac{1}{2}       \\
 \vspace{0.1cm}
\frac{1}{2\sqrt{5}}  & -\frac{1}{2}       & \frac{3}{2\sqrt{5}} & \frac{1}{2}       \\
\end{array}
\right) \cdot
\left(\begin{array}{c}
\vspace{0.13cm} g_{27}\\
\vspace{0.13cm}g_{\bar{10}}\\
\vspace{0.13cm}g_{8_s}\\
\vspace{0.13cm}g_{8_a}\\
 \end{array}
 \right)
 \label{eq:gSU3}
\end{equation}
\noindent
By using in the left hand side of Eq.~(\ref{eq:gSU3}) the values
obtained in Table~\ref{tab:poles5}
we get the results shown in Table~\ref{tab:gSU3}.

\begin{table}[h]
\begin{center}
\begin{tabular}{|c|c|c|}
\hline
               &  $1112-i64$     & $1216-i4$    \\	
\hline 
$g_{27}$       &  $-8+i94$       & $316-i21$  \\
$g_{\bar{10}}$ &  $-346+i15$     & $-406+i51$ \\
$g_{8_s}$      &  $4259-i2455$ & $-3783-i947$  \\ 
$g_{8_a}$      &  $2317-i928$    & $5208-i636$   \\
\hline
\end{tabular}
\caption{Couplings for $S=1$, $I=1/2$, in SU(3) base.}
\label{tab:gSU3}
\end{center}
\end{table}

As mentioned after Eq.~(\ref{eq:decomp}), the singlet
component of the $\phi$ and $\omega$ (see Eq.~(\ref{eq:phiw}))
does not lead to an interaction term, and this is already
assumed in Eq.~(\ref{eq:gSU3}) where only the octet component
of the $\phi$ is taken. We could as well have taken instead of
$g_{\phi K}$ in Eq.~(\ref{eq:gSU3}) the coupling
$-\sqrt{2}g_{\omega K}$ and in the exact $SU(3)$ limit this would
lead to the same results. When $SU(3)$ is broken, the use of
either coupling in the Eqs.~(\ref{eq:gSU3}) will give us an
idea about the uncertainties in this $SU(3)$ decomposition.

In Table~\ref{tab:gSU3} we can see that the couplings to the
$27-$plet and the decuplet are almost negligible in comparison to
the two octets, in fact, they are compatible with zero within the
uncertainties we can assume in our model.
This means that the two poles found are
essentially a mixing of the antisymmetric and symmetric octets.
The above reasoning is more than qualitative since it allows us
to quantify the weight of the $SU(3)$ components of the resonance
and therefore the mixing angle defined in Eq.~(\ref{eq:mixang}).
In order to obtain this mixing angle let us write  
Eq.~(\ref{eq:mixang}) in a more generic way:

\begin{equation}
|R>=a_{27}|27,R>+a_{\bar{10}}|\bar{10},R>
+a_{8_s}|8_s,R>+a_{8_a}|8_a,R>
\end{equation}
\noindent
where we  have used the short notation
$|27,R>\equiv|27,1,1/2,+1/2,R>$ and so on for the rest. The $a_j$
coefficients must satisfy $\sum |a_j|^2=1$.
To obtain the mixing angle $\theta$ we have to evaluate
$a_{8_s}$ and $a_{8_a}$ since, by definition, they are 
$a_{8_s}=\cos\theta$ and $a_{8_a}=\sin\theta$
for $|K_1(1270),1>$.

On the other hand, up to lowest
order in $SU(3)$ breaking, we have 

\begin{eqnarray}
\nonumber
g_{8_s}&\equiv&<R|t|8_s>\simeq a^*_{8_s}<8_s,R|t|8_s>
\equiv a^*_{8_s}t_{8_s}\qquad
a),\\
g_{8_a}&\equiv&<R|t|8_a>\simeq a^*_{8_a}<8_a,R|t|8_a>
\equiv a^*_{8_a} t_{8_a}\qquad
b).
\label{eq:thetag}
\end{eqnarray}
Therefore, had we know $t_{8_s}$, $t_{8_a}$, we would know 
$a_{8_s}$, $a_{8_a}$, by using  $g_{8_s}$, $g_{8_a}$, from
 Table~\ref{tab:gSU3}.
To evaluate $t_{8_s}$ and $t_{8_a}$ we can go to the $SU(3)$
limit, which by virtue of Eq.~(\ref{eq:CSU3})
should be the same. In this limit
 the two octet poles are in the same place.
The couplings $t_{8_s}$, $t_{8_a}$, are easily evaluated since
the matrix elements of the potential in the $SU(3)$ base are
already known in Eq.~(\ref{eq:CSU3}). Hence, by performing the
Bethe-Salpeter resummation in this base, we readily obtain the
poles corresponding to each $SU(3)$  irreducible representations
and the couplings. The amplitudes behave in the pole as
\begin{equation}
<8_s|T|8_s>=\frac{t_{8_s}^2}{s-s_p},
\end{equation}
\noindent
and analogously for $8_a$ and singlet, the only channels where
there are poles. We find $t_{8_s}=t_{8_a}=5568\mev$.

In the case of physical masses, $x=1$, using the same values
 for  $t_{8_s}$ and $t_{8_a}$ that we have obtained in the
 $SU(3)$ limit and the couplings $g_{8_s}$, $g_{8_a}$, of 
 Table~\ref{tab:gSU3},
  we obtain the following results:
for the pole $1112-i64\mev$ we get from Eq.~(\ref{eq:thetag}a)
$\theta\simeq 28^o$, and from Eq.~(\ref{eq:thetag}b) 
$\theta\simeq 27^o$.
Should we have 
used $-\sqrt{2}g_{\omega K}$ instead of $g_{\phi K}$ in
the calculations we would get $\theta\simeq 22^o$ from
Eq.~(\ref{eq:thetag}a) and $\theta\simeq 34^o$ from
Eq.~(\ref{eq:thetag}b). These discrepancies give an idea
of the uncertainties in our calculations.

For the pole $1216-i4\mev$, we get from Eq.~(\ref{eq:thetag}a)
$44^o$, and from Eq.~(\ref{eq:thetag}b) 
$20^o$. And with $g_{\omega K}$ $43^o$ and $11^o$ respectively,
which implies that we have a larger uncertainty in this case.
In fact, for the pole $1112-i64\mev$ we get
$\sum |a_j|^2=0.98$ by using $g_{\phi K}$
and $1.17$  by using $g_{\omega K}$,
and for the pole $1216-i4\mev$, $1.38$ and $1.43$. The
deviation of these numbers with respect to $1$ gives an idea
of the uncertainties.

I summary, we get a mixing angle of the order of
$30^o$ with an uncertainty of about $40$\%.
We should however refrain from making any association of 
the mixing angle found here with the mixing angle of
$K_{1A}$, $K_{1B}$, mentioned in the literature since this
latter one is used to produce the $K_1(1270)$ and $K_1(1400)$
resonances and here we are mixing $K_{1A}$, $K_{1B}$, to produce 
$|K_1(1270),1>$ and $|K_1(1270),2>$. \\

Along this work we have devoted some attention to uncertainties in
the theory. We would like to call the attention to another source
of uncertainties which would affect all channels. So far we have
used only the pion decay constant, $f=f_\pi=92\mev$, in the
Lagrangian. We could have used different values for $f_\pi$,
$f_K$, $f_\eta$. In order to estimate uncertainties from this
source we have proceeded as in \cite{kaon} and taken an average
$f=1.15f_\pi$. We find that all the poles obtained so far still
appear but the pole positions are somewhat changed. The
trajectories of the poles in the complex plane are essentially
the same but with real parts shifted about $50\mev$ to higher
energies (using the same subtraction constant). The imaginary
parts are accordingly increased since there is now more phase
space. The most significant changes in the imaginary part of
the pole positions are for the poles associated to the
$h_1(1170)$ ($100$\% increase), $a_1(1260)$ ($40$\% increase)
and $K_1(1270;1)$ ($50$\% increase)  and small changes in the
rest. These increases are mostly tied to the increased phase
space. However, we have so far evaluated the partial decay widths
of the resonances, using the couplings obtained and the physical
masses of the resonances. We have checked that the couplings of
the resonances barely change with the use of the new $f$ constant
(less than $10$\% change in general and less than $5$\% in the
dominant channels) and hence, within the uncertainties discussed
along the work, the results and the conclusions are unchanged
from this new source of uncertainty.

The overall conclusion of all these tests is that the existence
of the poles and their basic properties are very solid and not
contingent to the difference sources of uncertainties discussed
along the paper.

\section{Conclusions}

In this work we have done a systematic search for possible
$J^{PC}=1^{++},1^{+-}$, dynamically generated resonances through
the interaction of vector and pseudoscalar mesons. The starting
point has been a chiral Lagrangian which, by expanding up to two
pseudoscalar fields, leads to a Weinberg-Tomozawa term which
accounts for the
two-vectors and two-pseudoscalar meson interaction. From this
Lagrangian we have argued, by going to the $SU(3)$ limit that
from the interaction of the two octets we could expect
attraction for one singlet and two octets. After that, we have
implemented unitarity in coupled channels to account for the
resummation of $VP$ loops in $L=0$. This resummation has been
accounted for by means of a Bethe-Salpeter like equation in
coupled channels, where some subtleties in the evaluation of the
loop functions  coming from the use of vector mesons have been
discussed. The regularization of the loops has been done  with
dimensional regularization by means of a subtraction constant
fixed to agree with the numerical result obtained performing the
integration with a cutoff of natural size. We also compare our
method with another one where the loop functions are fixed to
zero when $\sqrt{s}$ is equal to the vector meson mass. This
served to have an idea of uncertainties in the theory.

We have looked for poles of the scattering amplitudes in the
second Riemann sheet. In the $SU(3)$ symmetric case, considered
by taking equal masses for the vectors and equal masses for the
pseudoscalars, we found two poles in the same position 
corresponding to two degenerate octets and one pole
corresponding to the singlet. As $SU(3)$ symmetry is gradually
broken, the two degenerate poles split apart in trajectories
(different for each $(S,I)$ channel) ending in the physical
situation, when the physical masses are used.

By evaluating the residues of the amplitudes in the poles, we
have obtained the couplings of the axial-vector dynamically
generated resonances to each $VP$ channel, which has allowed us
to evaluate the different partial decay widths. In view of the
information supplied by the pole positions, couplings and
partial widths, we have done  a correspondence of the poles
found to the $h_1(1170)$, $h_1(1380)$, $f_1(1285)$, $b_1(1235)$,
$a_1(1260)$ and $K_1(1270)$. For this latter case we found
actually two poles coupled strongly to $K^*\pi$ and $K\rho$
respectively which also differed appreciably in the width. We
suggested that different experiments give different weight to
each of these resonances and this could explain the
discrepancies in the widths obtained in different reactions.
This would also explain the correlation found between
experiments finding a dominance of the $\rho K$ decay (which
produce a small width) and those finding a dominance of the
$K^*\pi$ decay mode (which produce a large width). We also
evaluated the couplings of our two $K_1(1270)$ states, finding a
reasonable mixing of around $30^o$.

The only axial-vector resonances for which we do not find poles
are the $f_1(1420)$ and the $K_1(1400)$ for different reasons:
from the interaction of two octets one can only generate one
singlet and two octets, therefore there is no room for more
poles apart from those discussed above. Actually we only found a
pole with the $f_1$ quantum numbers which  suits better to the
$f_1(1285)$. In the two poles that we find in the $S=1$,
$I=1/2$, channel there are no clues to identify one of them to
the $K_1(1400)$ resonance. 

The conclusions reached in this paper about the dynamical nature
of these axial-vector mesons should have experimental
repercussions in the sense that, with the information obtained in
the present work, one can make predictions for production of
these resonances in different reactions, which are amenable of
experimental search. This has been the case for other
dynamically generated resonances and we hope the present paper
encourages work in this direction.

\section*{Acknowledgments}

Two of us, L.~R. and J.~S., acknowledge support from the
Ministerio de Educaci\'on y Ciencia. 
This work is partly supported by DGICYT contract number
BFM2003-00856, and the E.U. EURIDICE network contract no.
HPRN-CT-2002-00311. This research is part of the EU
Integrated Infrastructure Initiative Hadron Physics Project
under contract number RII3-CT-2004-506078.\\

{\Large{\bf Appendix~I: S-wave and on-shell $VP\to V'P'$ tree
level amplitude}}\\

Let us evaluate the s-wave projection of the tree level amplitude
for the process $V(q)P(p)\to V(q')P(p')$.

After performing the $SU(3)$ trace in Eq.~\ref{eq:L} we get
the general expression
\be
{\cal L}=-\frac{1}{4f^2}C_{ij}(
 \partial^\nu V_\mu \partial_\nu P V'^\mu P'
-\partial^\nu V_\mu P V'^\mu \partial_\nu P'
-V_\mu \partial_\nu P \partial^\nu V'^\mu P'
+V_\mu P \partial^\nu V'^\mu \partial_\nu P'),
\label{eq:Lexpand}
\ee
\noindent
where now $V_\mu$, $V'_\mu$, $P$ and $P'$ are the meson fields,
(not the $SU(3)$ matrices of mesons).
Eq.~(\ref{eq:Lexpand}) leads to the following amplitude:
\be
t_{ij}=-\frac{\epsilon\cdot\epsilon'}{4f^2}
C_{ij}(p+p')(q+q')=
-\frac{\epsilon\cdot\epsilon'}{4f^2}C_{ij}(s-u),
\ee
where $s=(p+q)^2=(p'+q')^2$ and $u=(p'-q)^2=(q'-p)^2$ are the
usual Mandelstam kinematical variables.

The partial wave expansion of the amplitude can be written as
\be
T=\sum (2l+1)f_l(s)P_l(x),
\ee
\noindent
with $x\equiv\cos\theta$, $\theta$ the center of mass scattering
angle and $P_{l}$ are the Legendre polynomials.

Hence, the s-wave projection of the
 scattering amplitude is
\be
f_{l=0}(s)=\frac{1}{2}\int_{-1}^{1}T(s,t(x'),u(x'))P_{l=0}(x')dx'
\ee
 The $l=0$ partial
wave is what we will call the potential $V_{ij}$.

Expressing $u$ in terms of $x$ and taking the momenta
on-shell, we have $u=m'^2+M^2-2E(p')E(q)-2|\vec
p'||\vec q|x$, and
 the term proportional to $x$ vanishes when
performing the integration, and thus the
s-wave projection of the amplitude reads
\be
V_{ij}(s)=-\frac{\epsilon\cdot\epsilon'}{4f^2}
C_{ij}(s-m'^2-M^2+2E(p')E(q)),
\ee
\noindent
with
\ba
E(p')=\frac{1}{2\sqrt{s}}(s-M'^2+m'^2)\qquad;\qquad
E(q)=\frac{1}{2\sqrt{s}}(s+M^2-m^2).
\ea

Altogether, the tree level on-shell and s-wave amplitude is
\be
V_{ij}(s)=-\frac{\epsilon\cdot\epsilon'}{8f^2} C_{ij}
\left[3s-(M^2+m^2+M'^2+m'^2)
-\frac{1}{s}(M^2-m^2)(M'^2-m'^2)\right].
\ee \\

{\Large{\bf Appendix~II: Vector mesons polarization in the
resummation of the loops}}\\

In this Appendix we are going to explain some subtleties about
the treatment of the polarization vectors of the vector mesons
in the resummation of the loops in the Bethe-Salpeter equation.

\begin{figure}[h]
\begin{center}
\includegraphics[width=0.3\textwidth]{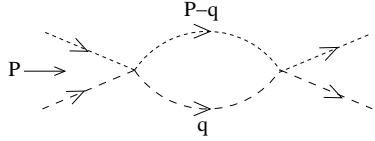}
\caption{Vector-Pseudoscalar loop.}
\label{fig:loop1}
\end{center}
\end{figure}

First let us show that, in order to look for poles, we can deal
only with the transverse vector-meson polarization modes.
Let us consider the case with only one channel (the
generalization to coupled channels is straightforward). Let us
consider the loop diagram of Fig.~\ref{fig:loop1}.

The Feynman rule is 

\be
t=V\epsilon_{\mu}i\, \int \frac{d^4 q}{(2 \pi)^4} \,
\frac{1}{(P-q)^2 - m^2 + i \epsilon} \,
 \frac{1}{q^2 - M^2 + i\epsilon}
 \left(-g^{\mu\nu}+\frac{q^\mu q^\nu}{M^2}\right)
 V\epsilon'_\nu,
 \label{eq:loop2}
\ee
\noindent
where we have taken for the tree level
$t\equiv V\epsilon\cdot\epsilon'$.

Note that we have factorized on-shell the $V$ functions 
which, as shown in \cite{ollernpa} using the Bethe-Salpeter
equation or  in \cite{ollerND} using the $N/D$ method, leads to a
well defined renormalization scheme. 

Since the integral only
depends on an external momentum $P$, the result of the integral
must be of the form
\be
b'g^{\mu\nu}+c'P^\mu P^\nu,
\ee
which can be written as
\be
V(bB^{\mu\nu}+cC^{\mu\nu})
\ee
with $B^{\mu\nu}\equiv g^{\mu\nu}-\frac{P^\mu P^\nu}{P^2}$
and $C^{\mu\nu}\equiv\frac{P^\mu P^\nu}{P^2}$, where we have
factorized one of the vertices, $V$.
The $B$ and $C$ tensors are idempotent and orthogonal in the
sense that
\ba
\nonumber B^{\mu\nu}B_\nu\,^\lambda=B^{\mu\lambda}, \\
\nonumber C^{\mu\nu}C_\nu\,^\lambda=C^{\mu\lambda}, \\
B^{\mu\nu}C_\nu\,^\lambda=0.
\ea
Therefore in the iteration of the loops we have two independent
series that sum up to

\be
\frac{Vb}{1-b}\left(g^{\mu\nu}-\frac{P^\mu P^\nu}{P^2}\right)
+\frac{Vc}{1-c}\frac{P^\mu P^\nu}{P^2}.
\label{eq:ab}
\ee
In the center of mass frame we have
$P^\mu=(\sqrt{s},0,0,0)$.
The transverse vector mesons (which have only
space components) only contribute to the left term of
Eq.~(\ref{eq:ab}), while the longitudinal vector mesons
 (which has
also time component in the polarization vectors) contribute to
both terms of Eq.~(\ref{eq:ab}).

Therefore, when looking for poles of the $T$ matrix, since if
there are poles they should exist for all modes and the second
term of Eq.~(\ref{eq:ab}) does not contribute for transverse
modes, the poles can only be in $Vb/(1-b)$ which is common for all
modes.
In practice we find that $c$ is one order of magnitude smaller
than $b$ and of opposite sign and the second term of 
Eq.~(\ref{eq:ab}) leads indeed to no poles.

Therefore, in order to look for poles we can only consider
transverse polarization vectors.
In this case the tree level $VP$ potential can be written as
$t=V\epsilon\cdot\epsilon'=-V\vec{\epsilon}\cdot\vec{\epsilon}\,'$,
and Eq.~(\ref{eq:loop2}) reads

\be
t=-V\epsilon_{i}\,i \int \frac{d^4 q}{(2 \pi)^4} \,
\frac{1}{(P-q)^2 - m^2 + i \epsilon} \,
 \frac{1}{q^2 - M^2 + i\epsilon}
 \left(-g_{ij}+\frac{q_i q_j}{M^2}\right)(-V)\epsilon'_j.
 \label{eq:loop3}
\ee

The term $q_i q_j$ can be replaced by
$1/3\vec{q}\,^2\delta_{ij}$. In addition, as shown in
Ref.~\cite{cabrera}, the $\vec{q}\,^2$ term can be taken
on-shell ($\vec{q}\,^2_{ON}=1/(4s)\lambda(s,M^2,m^2)$) since the
off-shell part  $\vec{q}\,^2-\vec{q}\,^2_{ON}$ can be reabsorbed
into the renormalization of the couplings. Therefore
Eq.~(\ref{eq:loop3}) can be expressed as,
\be
t=V^2\epsilon_i\epsilon_i
G(1+\frac{1}{3}\frac{\vec{q}\,^2_{ON}}{M^2}),
\ee
with
\be
G(\sqrt{s})= i \, \int \frac{d^4 q}{(2 \pi)^4} \,
\frac{1}{(P-q)^2 - m^2 + i \epsilon} \,
 \frac{1}{q^2 - M^2 + i
\epsilon}.
\ee
Therefore the series of Fig.~\ref{fig:bethe} would give
\be
T=-V\epsilon_i\epsilon_i
\ + \ (-V)^2\epsilon_i\epsilon_i
G(1+\frac{1}{3}\frac{\vec{q}\,^2_{ON}}{M^2})
\ + \ (-V)^3\epsilon_i\epsilon_i
G^2(1+\frac{1}{3}\frac{\vec{q}\,^2_{ON}}{M^2})^2+...
\ee
\noindent 
which sums up to
\be
T=\frac{-V}{1+VG(1+\frac{1}{3}\frac{\vec{q}\,^2_{ON}}{M^2})}
\,\vec{\epsilon}\cdot\vec{\epsilon}\,'
\label{eq:Tkk}
\ee

Eq.~(\ref{eq:Tkk}) can be easily generalized to more than one
channel giving a Bethe-Salpeter like coupled channel equation

\begin{equation}
T=[1+V\hat{G}]^{-1}(-V)
\,\vec{\epsilon}\cdot\vec{\epsilon}\,',
\label{eq:Tbetheapp}
\end{equation}
\noindent
where $\hat{G}=G(1+\frac{1}{3}\frac{q^2_l}{M_l^2})$
is a diagonal matrix with the $l-$th element, $G_l$, being
the two meson loop function
containing a vector and a pseudoscalar meson.
The term $\frac{1}{3}\frac{q^2_l}{M_l^2}$ is small and has no
much repercussion
in the final results.
By comparing Eq.~(\ref{eq:Tbetheapp}) and Eq.~(\ref{eq:ab}) we
can see that $b=-V\hat{G}$.

\end{document}